%% file: main.tex
\pdfoutput=1
\documentclass[12pt,a4paper]{article}
\usepackage{ifthen} % for conditional statements
\newboolean{pdflatex}
\setboolean{pdflatex}{true} % False for eps figures 

\newboolean{articletitles}
\setboolean{articletitles}{true} % False removes titles in references

\newboolean{uprightparticles}
\setboolean{uprightparticles}{false} %True for upright particle symbols

\def\paperauthors{LHCb collaboration} % Leave as is for PAPER, CONF and FIGURE
\def\paperasciititle{First measurement of b-jet mass with and without grooming} % Set ASCII title here !! MAKE sure it's only ASCII characters !! 
\def\papertitle{First measurement of $b$-jet mass with and without grooming} % Latex formatted title
\def\paperkeywords{{High Energy Physics}, {LHCb}} % Comma separated list
\def\papercopyright{\the\year\ CERN for the benefit of the LHCb collaboration} % new since 9/Apr/2018
\def\paperlicence{CC BY 4.0 licence}
\def\paperlicenceurl{https://creativecommons.org/licenses/by/4.0/}

\newif\ifEnableSectionTOCLinks
\EnableSectionTOCLinksfalse % deactivated

\input{preamble}

\usepackage{xspace}
\input{commands}

\usepackage{placeins}

\usepackage{afterpage}

\begin{document}

%%%%%%%%%%%%%%%%%%%%%%%%%
%%%%% Title     %%%%%%%%%
%%%%%%%%%%%%%%%%%%%%%%%%%
\renewcommand{\thefootnote}{\fnsymbol{footnote}}
\setcounter{footnote}{1}

% %%%%%%% CHOOSE TITLE PAGE--------
%\onecolumn
\input{title-LHCb-PAPER}

%\twocolumn
% %%%%%%%%%%%%% ---------

\renewcommand{\thefootnote}{\arabic{footnote}}
\setcounter{footnote}{0}

%%%%%%%%%%%%%%%%%%%%%%%%%%%%%%%%
%%%%%  Table of Content   %%%%%%
%%%%%%%%%%%%%%%%%%%%%%%%%%%%%%%%
%%%% Uncomment if desired
%\tableofcontents

\cleardoublepage

%%%%%%%%%%%%%%%%%%%%%%%%%
%%%%% Main text %%%%%%%%%
%%%%%%%%%%%%%%%%%%%%%%%%%

\pagestyle{plain} % restore page numbers for the main text
\setcounter{page}{1}
\pagenumbering{arabic}

%% Uncomment during review phase. 
%% Comment before a final submission.
%\linenumbers

%% This is the main body
\input{body}

% Comment this in for paper drafts; do not include this in analysis note, conference and figure reports
\input{acknowledgements}

\FloatBarrier

% This should be taken out in the final paper
\input{supplementary-app}

\FloatBarrier

\addcontentsline{toc}{section}{References}
%\setboolean{inbibliography}{true}
\bibliographystyle{LHCb}
\bibliography{bibliography,standard,LHCb-PAPER,LHCb-CONF,LHCb-DP,LHCb-TDR}

\newpage
\input{Authorship_LHCb-PAPER-2025-009}

\end{document}

%% file: preamble.tex
\usepackage[top=1in, bottom=1.25in, left=1in, right=1in]{geometry}

\usepackage{microtype}
\usepackage{lineno}  % for line numbering during review
\usepackage{xspace} % To avoid problems with missing or double spaces after
\usepackage{caption} %these three command get the figure and table captions automatically small

\usepackage{graphicx}  % to include figures (can also use other packages)
\usepackage{color}
\usepackage{colortbl}
\graphicspath{{./figs/}} % Make Latex search fig subdir for figures
\usepackage{amsmath} % Adds a large collection of math symbols
\usepackage{amssymb}
\usepackage{amsfonts}
\usepackage{upgreek} % Adds in support for greek letters in roman typeset

\newcommand*\patchAmsMathEnvironmentForLineno[1]{%
\expandafter\let\csname old#1\expandafter\endcsname\csname #1\endcsname
\expandafter\let\csname oldend#1\expandafter\endcsname\csname
end#1\endcsname
 \renewenvironment{#1}%
   {\linenomath\csname old#1\endcsname}%
   {\csname oldend#1\endcsname\endlinenomath}%
}
\newcommand*\patchBothAmsMathEnvironmentsForLineno[1]{%
  \patchAmsMathEnvironmentForLineno{#1}%
  \patchAmsMathEnvironmentForLineno{#1*}%
}
\AtBeginDocument{%
\patchBothAmsMathEnvironmentsForLineno{equation}%
\patchBothAmsMathEnvironmentsForLineno{align}%
\patchBothAmsMathEnvironmentsForLineno{flalign}%
\patchBothAmsMathEnvironmentsForLineno{alignat}%
\patchBothAmsMathEnvironmentsForLineno{gather}%
\patchBothAmsMathEnvironmentsForLineno{multline}%
\patchBothAmsMathEnvironmentsForLineno{eqnarray}%
}

\usepackage[pdftex,
            pdfauthor={\paperauthors},
            pdftitle={\paperasciititle},
            pdfkeywords={\paperkeywords}]{hyperref}
\usepackage{hyperxmp}
\hypersetup{
    pdfcopyright={Copyright (C) \papercopyright},
    pdflicenseurl={\paperlicenceurl}
}
\usepackage[colorinlistoftodos,textsize=scriptsize]{todonotes}

\usepackage[bottom,flushmargin,hang,multiple]{footmisc}

\usepackage[all]{hypcap} % Internal hyperlinks to floats.

\input{lhcb-symbols-def} % Add in the predefined LHCb symbols

\hypersetup{
  colorlinks   = true, %Colours links instead of ugly boxes
  urlcolor     = blue, %Colour for external hyperlinks
  linkcolor    = blue, %Colour of internal links
  citecolor    = red   %Colour of citations
}

\ifEnableSectionTOCLinks
    \usepackage[explicit]{titlesec} % to change headings
    
    \let\oldcontentsline\contentsline
    \renewcommand

    \titleformat{\section}{\normalfont\Large\bf}{\hyperlink{tocsection.\thesection}{{\thesection} \parbox[t]{\dimexpr\textwidth-1pc}{#1}}}{1pc}{}

    \titleformat{\subsection}{\normalfont\bf}{\hyperlink{tocsubsection.\thesubsection}{{\thesubsection} \parbox[t]{\dimexpr\textwidth-1pc}{#1}}}{1pc}{}

    \titleformat{name=\section,numberless}[display]{}{}{0pt}{\normalfont\Huge\bfseries #1}
\fi

\usepackage{cite} % Allows for ranges in citations
\usepackage{mciteplus}

%% file: lhcb-symbols-def.tex
\usepackage{xspace} 
\usepackage{upgreek}

\def\lhcb   {\mbox{LHCb}\xspace}

\def\cern {\mbox{CERN}\xspace}

\def\rich   {RICH\xspace}

\def\MagUp {\mbox{\em Mag\kern -0.05em Up}\xspace}

\ifthenelse{\boolean{uprightparticles}}%
{

 \def\Pmu         {\ensuremath{\upmu}\xspace}

 \def\Ppi         {\ensuremath{\uppi}\xspace}

 \def\Ppsi        {\ensuremath{\uppsi}\xspace}

 \def\PDelta      {\ensuremath{\Delta}\xspace}                 
 \def\PXi         {\ensuremath{\Xi}\xspace}                 
 \def\PLambda     {\ensuremath{\Lambda}\xspace}                 
 \def\PSigma      {\ensuremath{\Sigma}\xspace}                 
 \def\POmega      {\ensuremath{\Omega}\xspace}                 
 \def\PUpsilon    {\ensuremath{\Upsilon}\xspace}
 \let\oldPi\Pi
 \def\PPi         {\ensuremath{\oldPi}\xspace}

 \def\PB      {\ensuremath{\mathrm{B}}\xspace}                 
 \def\PD      {\ensuremath{\mathrm{D}}\xspace}                 
 \def\PJ      {\ensuremath{\mathrm{J}}\xspace}                 
 \def\PK      {\ensuremath{\mathrm{K}}\xspace}                 
 \def\PZ      {\ensuremath{\mathrm{Z}}\xspace}                 
 \def\Pb      {\ensuremath{\mathrm{b}}\xspace}                 
 \def\Pc      {\ensuremath{\mathrm{c}}\xspace}

 \def\Ps      {\ensuremath{\mathrm{s}}\xspace}

 \def\thebaroffset{0.0em}
}
{

 \def\Pmu         {\ensuremath{\mu}\xspace}

 \def\Ppi         {\ensuremath{\pi}\xspace}

 \def\Ppsi        {\ensuremath{\psi}\xspace}                 
                  
 \mathchardef\PDelta="7101
 \mathchardef\PXi="7104
 \mathchardef\PLambda="7103
 \mathchardef\PSigma="7106
 \mathchardef\POmega="710A
 \mathchardef\PUpsilon="7107
 \mathchardef\PPi="7105
 \def\PB      {\ensuremath{B}\xspace}                 
 \def\PD      {\ensuremath{D}\xspace}                 
 \def\PJ      {\ensuremath{J}\xspace}                 
 \def\PK      {\ensuremath{K}\xspace}                 
 \def\PZ      {\ensuremath{Z}\xspace}                 
 \def\Pb      {\ensuremath{b}\xspace}                 
 \def\Pc      {\ensuremath{c}\xspace}

 \def\Ps      {\ensuremath{s}\xspace}

 \def\thebaroffset{0.18em}
}
\newcommand{\offsetoverline}[2][\thebaroffset]{\kern #1\overline{\kern -#1 #2}}%

\makeatletter
\ifcase \@ptsize \relax% 10pt
  \newcommand{\miniscule}{\@setfontsize\miniscule{4}{5}}% \tiny: 5/6
\or% 11pt
  \newcommand{\miniscule}{\@setfontsize\miniscule{5}{6}}% \tiny: 6/7
\or% 12pt
  \newcommand{\miniscule}{\@setfontsize\miniscule{5}{6}}% \tiny: 6/7
\fi
\makeatother

\DeclareRobustCommand{\optbar}[1]{\shortstack{{\miniscule (\rule[.5ex]{1.25em}{.18mm})}
  \\ [-.7ex] $#1$}}

\def\mup        {{\ensuremath{\Pmu^+}}\xspace}
\def\mun        {{\ensuremath{\Pmu^-}}\xspace} % muon negative (\mum is taken)
\def\mupm       {{\ensuremath{\Pmu^\pm}}\xspace}

\def\Z      {{\ensuremath{\PZ}}\xspace}

\def\squark    {{\ensuremath{\Ps}}\xspace}

\def\cquark    {{\ensuremath{\Pc}}\xspace}

\def\bquark    {{\ensuremath{\Pb}}\xspace}

\def\pion   {{\ensuremath{\Ppi}}\xspace}
\def\piz    {{\ensuremath{\pion^0}}\xspace}

\def\pipm   {{\ensuremath{\pion^\pm}}\xspace}

\def\kaon    {{\ensuremath{\PK}}\xspace}
\def\KorKbar {\kern \thebaroffset\optbar{\kern -\thebaroffset \PK}{}\xspace}

\def\Kpm     {{\ensuremath{\kaon^\pm}}\xspace}

\def\D       {{\ensuremath{\PD}}\xspace}

\def\DorDbar {\kern \thebaroffset\optbar{\kern -\thebaroffset \PD}\xspace}

\def\Dp      {{\ensuremath{\D^+}}\xspace}
\def\Dm      {{\ensuremath{\D^-}}\xspace}

\def\DpDm    {\ensuremath{\Dp {\kern -0.16em \Dm}}\xspace}

\def\B       {{\ensuremath{\PB}}\xspace}

\def\BorBbar {\kern \thebaroffset\optbar{\kern -\thebaroffset \PB}\xspace}
\def\Bz      {{\ensuremath{\B^0}}\xspace}

\def\Bd      {{\ensuremath{\B^0}}\xspace}

\def\BdorBdbar {\kern \thebaroffset\optbar{\kern -\thebaroffset \Bd}\xspace}

\def\Bpm     {{\ensuremath{\B^\pm}}\xspace}

\def\Bs      {{\ensuremath{\B^0_\squark}}\xspace}

\def\BsorBsbar {\kern \thebaroffset\optbar{\kern -\thebaroffset \Bs}\xspace}

\def\jpsi     {{\ensuremath{{\PJ\mskip -3mu/\mskip -2mu\Ppsi}}}\xspace}

\def\Y#1S{\ensuremath{\PUpsilon{(#1S)}}\xspace}

\def\LorLbar     {\kern \thebaroffset\optbar{\kern -\thebaroffset \PLambda}\xspace}

\def\to                 {\ensuremath{\rightarrow}\xspace}

\def\AT#1     {\ensuremath{A_{\mathrm{T}}^{#1}}\xspace}           % 2

\def\C#1      {\ensuremath{\mathcal{C}_{#1}}\xspace}                       % 9
\def\Cp#1     {\ensuremath{\mathcal{C}_{#1}^{'}}\xspace}                    % 7
\def\Ceff#1   {\ensuremath{\mathcal{C}_{#1}^{\mathrm{(eff)}}}\xspace}        % 9  
\def\Cpeff#1  {\ensuremath{\mathcal{C}_{#1}^{'\mathrm{(eff)}}}\xspace}       % 7
\def\Ope#1    {\ensuremath{\mathcal{O}_{#1}}\xspace}                       % 2
\def\Opep#1   {\ensuremath{\mathcal{O}_{#1}^{'}}\xspace}                    % 7

\newcommand{\aunit}[1]{\ensuremath{\text{\,#1}}}       
\newcommand{\tev}{\aunit{Te\kern -0.1em V}\xspace}
\newcommand{\gev}{\aunit{Ge\kern -0.1em V}\xspace}
\newcommand{\mev}{\aunit{Me\kern -0.1em V}\xspace}
\newcommand{\kev}{\aunit{ke\kern -0.1em V}\xspace}
\newcommand{\ev}{\aunit{e\kern -0.1em V}\xspace}
 
\newcommand{\mevc}{\ensuremath{\aunit{Me\kern -0.1em V\!/}c}\xspace}
\newcommand{\gevc}{\ensuremath{\aunit{Ge\kern -0.1em V\!/}c}\xspace}
\newcommand{\mevcc}{\ensuremath{\aunit{Me\kern -0.1em V\!/}c^2}\xspace}
\newcommand{\gevcc}{\ensuremath{\aunit{Ge\kern -0.1em V\!/}c^2}\xspace}
\def\gsim{{~\raise.15em\hbox{$>$}\kern-.85em
          \lower.35em\hbox{$\sim$}~}\xspace}
\def\lsim{{~\raise.15em\hbox{$<$}\kern-.85em
          \lower.35em\hbox{$\sim$}~}\xspace}

\def\pt         {\ensuremath{p_{\mathrm{T}}}\xspace}

\def\evtgen     {\mbox{\textsc{EvtGen}}\xspace}

\def\geant      {\mbox{\textsc{Geant4}}\xspace}

\def\photos     {\mbox{\textsc{Photos}}\xspace}

\def\pythia     {\mbox{\textsc{Pythia}}\xspace}

\def\tell1  {TELL1\xspace}
\def\ukl1   {UKL1\xspace}

\newcommand{\lhcborcid}[1]{\href{https://orcid.org/#1}{\hspace*{0.1em}\raisebox{-0.45ex}{\includegraphics[width=1em]{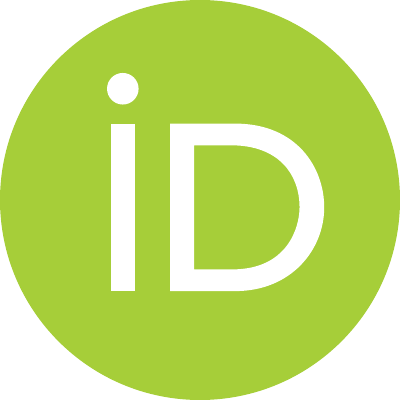}}}}

%% file: commands.tex
\newcommand{\pp}           {\textit{pp}\xspace}

\newcommand{\s}            {\ensuremath{\sqrt{s}}\xspace}

\newcommand{\nineH}        {$\sqrt{s}~=~0.9$~Te\kern-.1emV\xspace}
\newcommand{\seven}        {$\sqrt{s}~=~7$~Te\kern-.1emV\xspace}
\newcommand{\twoH}         {$\sqrt{s}~=~0.2$~Te\kern-.1emV\xspace}
\newcommand{\twosevensix}  {$\sqrt{s}~=~2.76$~Te\kern-.1emV\xspace}
\newcommand{\five}         {$\sqrt{s}~=~5.02$~Te\kern-.1emV\xspace}
\newcommand{\twosevensixnn}{$\sqrt{s_{\mathrm{NN}}}~=~2.76$~Te\kern-.1emV\xspace}
\newcommand{\fivenn}       {$\sqrt{s_{\mathrm{NN}}}~=~5.02$~Te\kern-.1emV\xspace}

\newcommand{\GeVc}         {Ge\kern-.1emV/$c$\xspace}
\newcommand{\MeVc}         {Me\kern-.1emV/$c$\xspace}
\newcommand{\TeV}          {Te\kern-.1emV\xspace}
\newcommand{\GeV}          {Ge\kern-.1emV\xspace}
\newcommand{\MeV}          {Me\kern-.1emV\xspace}
\newcommand{\GeVmass}      {Ge\kern-.2emV/$c^2$\xspace}
\newcommand{\MeVmass}      {Me\kern-.2emV/$c^2$\xspace}
\newcommand{\kt}           {\ensuremath{k_\mathrm{T}}\xspace}
\newcommand{\pTjet}        {\ensuremath{p_{\mathrm{T,jet}}}\xspace}
\newcommand{\pTHF}         {\ensuremath{p_{\mathrm{T,HF}}}\xspace}
\newcommand{\mjet}         {\ensuremath{m_\mathrm{jet}}\xspace}
\newcommand{\mjetsd}       {\ensuremath{m_\mathrm{jet,gr}}\xspace}
\newcommand{\zcut}         {\ensuremath{z_\mathrm{cut}}\xspace}

%% file: title-LHCb-PAPER.tex
% ===============================================================================
% Purpose: LHCb-PAPER journal paper title page template
% Author: 
% Created on: 2010-09-25
% ===============================================================================

%%%%%%%%%%%%%%%%%%%%%%%%%
%%%%%  TITLE PAGE  %%%%%%
%%%%%%%%%%%%%%%%%%%%%%%%%
\begin{titlepage}
\pagenumbering{roman}

% Header ---------------------------------------------------
\vspace*{-1.5cm}
\centerline{\large EUROPEAN ORGANIZATION FOR NUCLEAR RESEARCH (CERN)}
\vspace*{1.5cm}
\noindent
\begin{tabular*}{\linewidth}{lc@{\extracolsep{\fill}}r@{\extracolsep{0pt}}}
\ifthenelse{\boolean{pdflatex}}% Logo format choice
{\vspace*{-1.5cm}\mbox{\!\!\!\includegraphics[width=.14\textwidth]{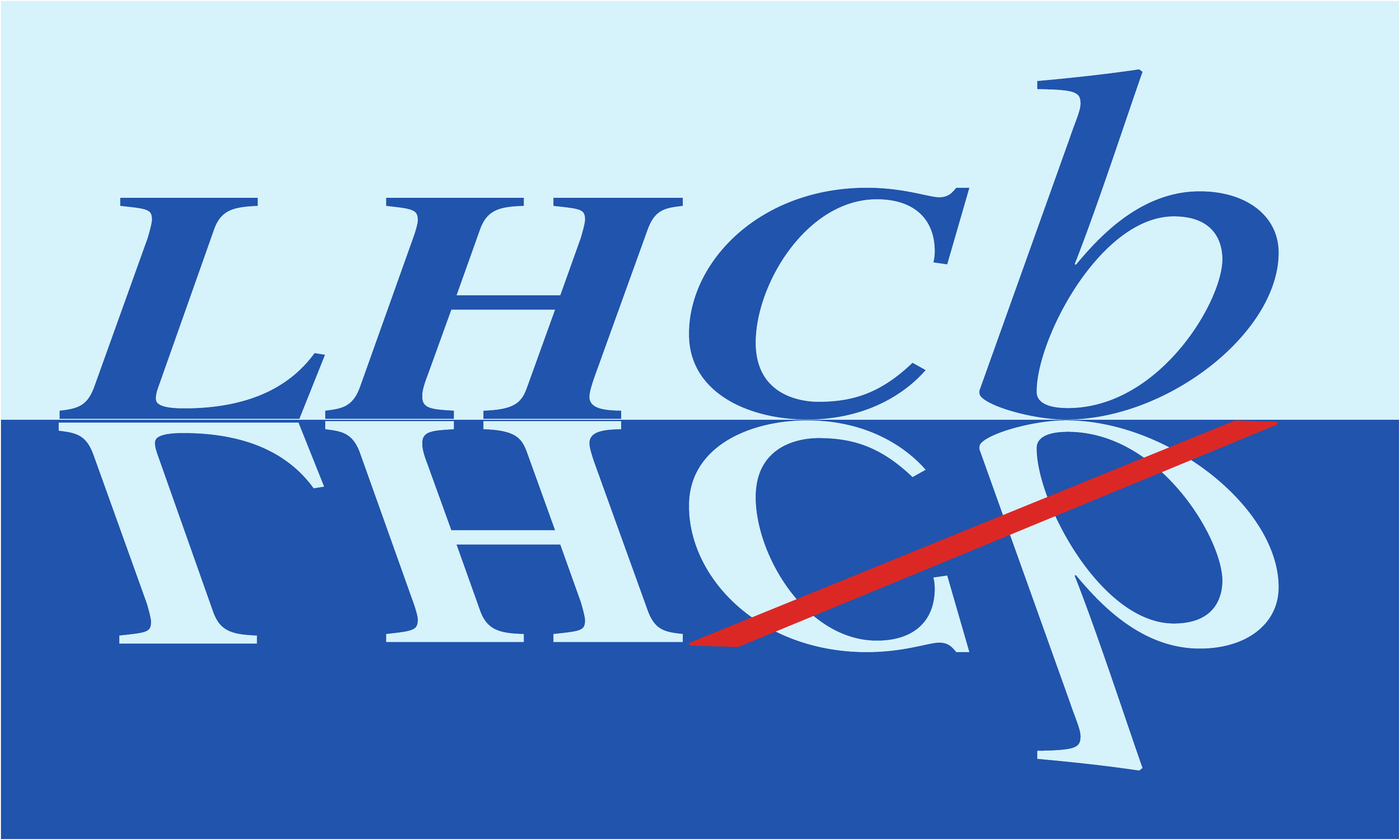}} & &}%
{\vspace*{-1.2cm}\mbox{\!\!\!\includegraphics[width=.12\textwidth]{figs/lhcb-logo.eps}} & &}%
\\
 & & CERN-EP-2025-097 \\  % ID 
 & & LHCb-PAPER-2025-009 \\  % ID 
 & & November 3, 2025 \\ % Date - Can use \today or hardwire, e.g.: 23 March 2010
 & & \\
% not in paper \hline
\end{tabular*}

\vspace*{3.0cm}

% Title --------------------------------------------------
{\normalfont\bfseries\boldmath\huge
\begin{center}
% DO NOT EDIT HERE. Instead edit macro in main.tex to keep metadata correct
  \papertitle 
\end{center}
}

\vspace*{1.0cm}

% Authors -------------------------------------------------
\begin{center}
%In the footnote, replace 'paper' by 'Letter' in case of submission to PRL or PLB 
% Edit macro in main.tex to keep metadata correct
\paperauthors\footnote{Authors are listed at the end of this paper.}
\end{center}

\vspace{\fill}

% Abstract -----------------------------------------------
\begin{abstract}
\noindent
The LHCb collaboration presents a novel suite of heavy-flavour jet substructure measurements at forward rapidity in proton-proton collisions at a centre-of-mass energy of $\sqrt{s} = 13$\tev.
The jet mass is a perturbatively calculable probe of the virtuality of hard-scattered quarks and gluons, connecting small-distance quantum chromodynamics (QCD) with long-distance experimental measurement. The jet mass is dominated by nonperturbative corrections at small values, presenting an excellent test of QCD across a broad range of jet energies.
Measuring heavy-flavour jet mass with a theoretically unambiguous flavour definition for the first time probes the gluon splitting mechanism for heavy-flavour production and offers tests of perturbative QCD at new levels of theoretical precision. 
Utilising the soft drop jet-grooming technique to access the perturbative jet core further enhances constraints on first-principles theory. 
Measurements of the jet mass for jets containing fully reconstructed $B^\pm$ hadrons are reported with and without grooming.
The unique phase space instrumented by LHCb offers a different quark-gluon fraction than at midrapidity.
These results offer unparalleled tests of quark flavour and mass dependence in QCD and provide a baseline for future studies of heavy-flavour jet quenching in heavy-ion collisions.
\end{abstract}

\vspace*{1.0cm}

\begin{center}
Published in Phys. Lett. B869 (2025) 139854
\end{center}

\vspace{\fill}

{\footnotesize 
% Edit macro in main.tex to keep metadata correct
\centerline{\copyright~\papercopyright. \href{\paperlicenceurl}{\paperlicence}.}}
\vspace*{2mm}

\end{titlepage}

%%%%%%%%%%%%%%%%%%%%%%%%%%%%%%%%
%%%%%  EOD OF TITLE PAGE  %%%%%%
%%%%%%%%%%%%%%%%%%%%%%%%%%%%%%%%

%  empty page follows the title page ----
\newpage
\setcounter{page}{2}
\mbox{~}
%\newpage
%
%% Author List ----------------------------
%%  You need to get a new author list!
%\input{LHCb_authorlist.tex}
%
%The author list for journal publications is provided by the Membership Committee shortly after 'approval to go to paper' has been given.
%%It will be made available on the page
%%\verb!http://www.physik.uzh.ch/~strauman/forMemCo/LHCb-PAPER-XXXX-XXX/! .
%It will be sent to you by email shortly after a paper number has beens assigned.
%The author list should be included already at first circulation, 
%to allow new members of the collaboration to verify whether they have been included correctly.
%Occasionally a misspelled name is corrected or associated institutions become full members.
%In that case, a new author list will be sent to you.
%In case line numbering doesn't work well after including the authorlist, try moving the \verb!\bigskip! after the last author to a separate line.
%
%
%The authorship for Conference Reports should be ``The LHCb
%  collaboration'', with a footnote giving the name(s) of the contact
%  author(s), but without the full list of collaboration names.

%% file: body.tex
\section{Introduction}

Collimated sprays of particles called jets are abundantly produced in high-energy particle collisions at the Large Hadron Collider (LHC), via the hard scattering of partons (quarks and gluons) from colliding nuclei. 
Scattered partons carry away some off-mass-shell virtuality, \mbox{$Q \equiv \sqrt{E^2 - p^2}$}, from the four-momentum exchanged in the interaction, the value of which is fundamentally unknown to an observer. However, in the parton shower picture, virtuality is shed by the successive radiation of gluons, forming a jet~\cite{SJOSTRAND1982243, Webber:1986mc, Sch_licke_2004, Buckley_2011}. Studying the final-state jet substructure allows access to the radiation pattern of the scattered parton, and in turn probes its pre-shower kinematics~\cite{Substructure_intro, Substructure_experiment_review}.

Insomuch as the experimentally reconstructed jet is a proxy for the initiating parton, the jet mass provides a proxy for the parton's virtuality,
\begin{equation}
\mjet = \sqrt{E_\mathrm{jet}^2 - p_\mathrm{jet}^2} \sim Q_\mathrm{parton},
\end{equation}
where $E_\mathrm{jet}$ is the jet energy and $p_\mathrm{jet}$ is its total momentum. Measurements of \mjet as a function of jet transverse momentum \pTjet therefore probe the dependence of partonic virtuality versus the momentum scale of the initial scattering. It is useful to study \mjet normalised by the jet transverse momentum,
\mjet/\pTjet,
which is dimensionless\footnote{Natural units with $\hbar=c=1$ are used throughout.} and closely related to the widely studied jet angularities~\cite{Berger_2003_1, Berger_2003_2, Berger_2004, Larkoski_2014, Kang_2018}. Because higher-virtuality quarks shed their virtuality with more QCD radiation, jets which are broader and contain more gluon emissions will tend to have a higher mass than jets which are narrower and contain less radiation.

Jet substructure calculations are theoretically challenging due to the presence of large non-global logarithms (NGLs) which arise due to the local phase-space constraints of jet observables~\cite{Dasgupta:2001sh}. Such contributions can be eliminated, however, by utilising \emph{soft drop grooming} to remove soft, wide-angle radiation from the jet~\cite{Dasgupta:2013ihk, Soft_drop}. Fixed-order calculations of \mjetsd, the mass of jets groomed with soft drop, have recently been computed up to next-to-next-to-leading order (N$^2$LO) including resummation of large logarithms up to next-to-next-to-next-to-leading logarithmic accuracy (N$^3$LL)~\cite{NNLO_mjet_gr}, offering an opportunity to study QCD via jet substructure at unprecedented precision.

Jets initiated by a heavy-flavour (HF) quark, \cquark{} or \bquark{}, have also become increasingly accessible at the LHC. The substructure of HF jets is modified relative to light-flavour jets due to the QCD ``dead cone'' effect~\cite{Dokshitzer_1991}, a perturbatively well-described suppression of gluon radiation at small angles from the hard-scattered HF quark. Heavy-flavour jets also probe nonperturbative effects such as the HF component of the proton wavefunction~\cite{LHCb-PAPER-2021-029} and HF fragmentation functions~\cite{LHCB-PAPER-2016-064, LHCb-PAPER-2019-012,  ATLAS_bjet_FF, LHCb-PAPER-2022-013, ALICE_D0_FF, ALICE_Lc_FF, LHCb-PAPER-2024-021, SD_HF}.
However, theoretical interpretation can be limited by traditional jet flavour tagging, as solely requiring a specific HF hadron to be reconstructed inside a jet is not theoretically well-defined past NLO~\cite{BSZ_2006}. Reconstructing secondary vertices and using machine learning to tag the jet flavour is also dependent on simulation models, which limits perturbative accuracy.

Recently a variety of algorithms calculable beyond NLO have been proposed to tag jet flavour while maintaining traditional jet kinematics~\cite{Caletti:2022hnc, IFN, CMP, GHS}. One novel approach is the Winner-Take-All (WTA) flavour-tagging algorithm~\cite{WTA_tagging}, which is calculable to any perturbative order~\cite{Larkoski:2023upz}.
At forward rapidity, adding a WTA flavour requirement predominantly removes jets whose HF content arises from in-shower gluon splitting ($g \rightarrow b \bar{b}$)~\cite{Les_Houches_studies}. Gluon splitting contaminates samples of direct HF production; however, it has also been proposed as a probe for studying bulk QCD dynamics in heavy-ion collisions~\cite{Jasmine_g2cc, PhysRevLett.132.212301}. The difference in \mjet/\pTjet before and after WTA tagging therefore offers a novel baseline for jet quenching~\cite{Bjorken_quenching, Bjorken_QGP, Appel_quenching, 1990_quenching} via a specific NLO process, useful for probing mass-dependent momentum transfer and quantum coherence effects in quark-gluon plasma~\cite{Jasmine_g2cc, PhysRevLett.132.212301, TheBigPicture}.

Previous studies of \mjet have been performed by the ATLAS~\cite{ATLAS_mass_bb, ATLAS_gr_mass_1, ATLAS_gr_mass_2, ATLAS_mass_Z}, CMS~\cite{CMS_gr_mass, CMS_mass_diff, CMS_mass_dijet, CMS_mass_ttbar}, and ALICE~\cite{ALICE_mjet_2017, ALICE_ang_PbPb_Run2} experiments at the LHC, as well as the STAR~\cite{STAR_mjet_pp} experiment at RHIC, but there has been no measurement for jets explicitly tagged with \bquark flavour. This letter presents new studies of \mjet for \bquark jets at forward rapidity in proton-proton (\pp) collisions at a centre-of-mass energy of $\s = 13$\tev with the \lhcb detector. Studying \bquark jets offers enhanced sensitivity to HF mass effects as compared to \cquark-tagged jets due to the larger \bquark-quark mass.
Jets containing a \bquark{} quark are identified using \Bpm mesons explicitly reconstructed from the $\jpsi \left(\rightarrow \mup \mun \right) \Kpm$ final state.
Fully reconstructing the \Bpm meson enables the use of WTA flavour tagging and improves jet momentum resolution as compared to statistical flavour tagging methods.
The unique acceptance of \lhcb offers a large \bquark-quark cross-section with significant longitudinal boost, enabling precision identification of secondary vertices from the \Bpm meson decay.
Jets are considered both with/without soft drop grooming~\cite{Soft_drop}, and with/without WTA flavour tagging~\cite{WTA_tagging}, and results are compared to \pythia simulations, with direct QCD calculations from the theory community still in preparation.

%%%%%%%%%%%%%%%%%%%%%%%%%%%%%%%%%%%%%%%%%%%%%%%%%%%%%%%%%%%%%%%%%%%
\section{Detector and dataset}
\label{sec:detector}
 
The \lhcb detector~\cite{LHCb-DP-2008-001, LHCb-DP-2014-002} is a single-arm forward spectrometer covering the
pseudorapidity range $2 < \eta < 5$, designed for the study of particles containing \bquark
and \cquark quarks. The detector includes a high-precision tracking system consisting of a
silicon-strip vertex detector surrounding the \pp interaction region~\cite{LHCb-DP-2014-001}, a large-area silicon-strip
detector located upstream of a dipole magnet with a bending power of about $4{\mathrm{\,T\,m}}$, and
three stations of silicon-strip detectors and straw drift tubes placed downstream of the
magnet~\cite{LHCb-DP-2017-001}. The tracking system measures the momentum, $p$, of charged particles with a
relative uncertainty that varies from 0.5\% at low momentum to 1.0\% at 200\gev~\cite{LHCb-DP-2014-002}. The
minimum distance of a track to a primary \pp collision vertex, called the impact
parameter, is measured with a resolution of $(15 + 29/\pt)$ $\mu$m, where \pt is in units
of\gev. Different types of charged hadrons are distinguished using information from
two ring-imaging Cherenkov detectors (\rich)~\cite{LHCb-DP-2012-003}. Photons, electrons and hadrons are identified by a calorimeter system consisting of scintillating-pad and preshower detectors, an
electromagnetic and a hadronic calorimeter. Muons are identified by a system composed
of alternating layers of iron and multiwire proportional chambers~\cite{LHCB-DP-2012-002}.

The data sample used in this analysis corresponds to an integrated luminosity of \mbox{$5.4$ fb$^{-1}$} recorded in 2016--2018.
The online event selection is performed by a trigger~\cite{LHCb-DP-2012-004} which consists of a
hardware stage using information from the calorimeter and muon systems, followed by
a software stage which performs the full event and candidate \Bpm meson reconstruction. 
The hardware stage selects events with at least one $\mupm$ candidate with \mbox{$\pt(\mupm) > 1.35$--1.80\gev},
or at least one pair of oppositely charged muons with \mbox{$\sqrt{\pt(\mup)\pt(\mun)} > 1.3$--1.5\gev},
where the threshold varied during the data taking period~\cite{LHCb-DP-2019-001}. In the software stage, two
oppositely charged $\mupm$ candidates, each with $\pt(\mupm) > 0.5$\gev, are required to form a \jpsi
candidate with a mass within $\pm 100$\mev of the known value~\cite{PDG2024} and $\pt > 2\gev$.
Information from the RICH detectors is used to identify \Kpm mesons, which reduces contamination from the Cabibbo-suppressed decay $\Bpm \to \jpsi(\to \mup \mun) \pipm$.

Simulation is required to model the effects of the detector response and the imposed
selection requirements on the $\jpsi (\to \mup \mun) K^\pm$ final state. In the simulation, \pp collisions
are generated using \pythia 8.186~\cite{Sjostrand:2007gs} with a specific \lhcb configuration~\cite{LHCb-PROC-2010-056}. Decays
of unstable particles are described by \evtgen~\cite{Lange:2001uf}, in which final-state radiation is
generated using \photos~\cite{davidson2015photos}. The interaction of the generated particles with the detector,
and its response, are implemented using the \geant toolkit~\cite{Agostinelli:2002hh, *Allison:2006ve} as described in Ref.~\cite{LHCb-PROC-2011-006}. To improve the simulated response of the \rich detectors, particle identification (PID) variables are corrected using calibration data from $D^{*+} \to D^0\pi^+$ and $D^0 \to K^-\pi^+$ decays for $K^\pm$ and $\pi^\pm$ mesons, and $\Lambda^+_c \to pK^-\pi^+$ decays for protons~\cite{LHCb-DP-2018-001}. This correction adjusts the variables to match the data distributions, accounting for track kinematics.

%%%%%%%%%%%%%%%%%%%%%%%%%%%%%%%%%%%%%%%%%%%%%%%%%%%%%%%%%%%%%%%%%%%
\section{Event reconstruction and jet selection}

To minimise uncorrelated in-jet radiation from pileup, events are required to have only one reconstructed primary vertex. Candidate \jpsi mesons are constrained to the known \jpsi mass~\cite{PDG2024} and combined with identified \Kpm mesons to construct candidate \Bpm mesons. These candidates are required to have a mass within 130\mev (below) and 300\mev (above) the known \Bpm mass~\cite{PDG2024}. The asymmetry in the mass window removes the partially reconstructed background coming from \Bpm and \Bz decays with an additional unreconstructed \piz or \pipm meson.
The reconstructed \Bpm meson candidates are required to originate from the primary vertex, which improves the signal purity and the mass resolution. 

After replacing its decay products with the reconstructed \Bpm candidate, jets are reconstructed with the \textsc{FastJet} 3.4.1 package~\cite{cacciari2012fastjet} using the anti-$\kt$ algorithm~\cite{anti-kT} with jet resolution (radius) parameter $R=0.5$ and $E$-scheme recombination.\footnote{$E$-scheme recombination determines the reconstructed jet energy and momentum by adding the four-vectors of the jet constituents.}
Particles are assigned the mass~\cite{PDG2024} of their most likely identity, utilising the excellent PID capabilities of \lhcb~\cite{LHCb-DP-2021-004}.
Jets that contain the candidate \Bpm meson are selected.
A minimum jet transverse momentum cut of $\pTjet > 5$\gev is applied. The jet rapidity range is restricted to $2.5 < y_\mathrm{jet} < 4.0$ to ensure all particles inside the jet are within the \lhcb detector acceptance and have good reconstruction efficiency.

A sample of groomed jets is produced using the soft drop algorithm~\cite{Soft_drop}.
The jet is reclustered from its constituents into a tree-like data structure using the Cambridge/Aachen algorithm~\cite{C/A_alg_1, *C/A_alg_2}, which preferentially clusters together particles nearby in the rapidity ($y$) -- azimuth ($\varphi$) plane. This leads to an angular ordering of branches in the tree, reflecting the approximate angular ordering of emissions in QCD~\cite{WEBBER1984492, MARCHESINI19841, MARCHESINI1988461}. Each splitting is then declustered starting from the last (widest-angle) pair, and the relative \pt fraction of the softer branch, $z \equiv p_{\mathrm{T,soft}}/(p_{\mathrm{T,soft}}+p_{\mathrm{T,hard}})$, is calculated. If $z < \zcut \theta^\beta$, where $\theta \equiv (\sqrt{\Delta y ^2 + \Delta \varphi ^2}) / R$, and $\zcut$ and $\beta$ are user-defined grooming parameters, then the softer branch is dropped, and the procedure continues with the next splitting in the remaining branch.
If, however, $z > \zcut \theta^\beta$, then the procedure is concluded, with all remaining constituents defining the groomed jet.
If the jet has no remaining splittings after removing a branch, then the
jet is ``groomed away'' and is removed from the data sample. Because they have no splittings, all single-constituent jets are groomed away. Grooming parameters $\zcut=0.1$ and $\beta=0$ are used.

Both ungroomed and groomed jets are evaluated using WTA flavour tagging~\cite{WTA_tagging}. As with soft drop, the jet is reclustered from its constituent particles using the Cambridge/Aachen algorithm. The WTA axis~\cite{Bertolini_2014, Larkoski:2014uqa} is determined by evaluating each recombination in the \pt scheme, whereby the combined axis is set to be along the direction of the higher $\pt$ branch, and is assigned a magnitude equal to the $\pt$ sum of the two branches. This continues until all particles are recombined. The WTA flavour tag is then determined by requiring the candidate \Bpm meson to lie along the WTA axis.

The \bquark-jet sample contains contamination from combinatorial background which must be statistically subtracted from the \mjet/\pTjet distributions. To identify the fraction of signal and background contributions, the candidate \Bpm mass spectrum is fit
using parametric signal and background models as a function of \Bpm transverse momentum, \pTHF. The signal peak is modelled using two double-sided Crystal Ball (DSCB) functions~\cite{Skwarnicki:1986xj, ROOT_DSCB} with shared peak position and tail parameters, with the latter constrained from fits in simulation.
The combinatorial background is modelled by a second-order polynomial function. The distribution of misidentified $B^\pm$ candidates arising from the Cabibbo-suppressed \mbox{$\Bpm \to \jpsi(\to \mup \mun) \pipm$} decay is determined in simulation and is also modelled using a DSCB function, but this background is found to be negligible due to strong PID requirements on the \Kpm meson.

``Signal plus background'' (S+B) and ``pure background'' (PB) regions are then defined as a function of candidate \Bpm mass and \pTHF. The S+B region is defined underneath the signal mass peak, with \pTHF-dependent edges typically between 5.25 and 5.31\gev.
The fraction of combinatorial background which contaminates this region falls from 10\% for $\pTHF < 4$\gev to negligible values for $\pTHF > 35$\gev. Two PB regions are then defined using sidebands away from the signal region, typically below 5.22\gev and above 5.36\gev.
Distributions of \mjet/\pTjet are measured using candidate \bquark jets in each region. The PB distribution is scaled by the ratio of background in the S+B region versus the PB region, and then statistically subtracted from the S+B distributions as a function of \pTHF to yield pure signal distributions. See the supplementary material for an example mass fit distribution.

%%%%%%%%%%%%%%%%%%%%%%%%%%%%%%%%%%%%%%%%%%%%%%%%%%%%%%%%%%%%%%%%%%%
\section{Detector corrections}

Effects originating from the imperfect response of the \lhcb detector are corrected by applying a suite of detector corrections. In simulation, jets containing a \Bpm hadron are geometrically matched in $y$ and $\varphi$ before and after the \lhcb detector simulation to estimate such effects. The \mjet/\pTjet distribution, binned differentially in \pTHF and \pTjet, is first corrected for its imperfect purity by removing contributions which originate from fake \bquark jets (those that do not map to a true \bquark jet) or \bquark jets from outside the reported kinematic region. The signal purity rises steeply from around 40--50\% (with WTA tagging) or 55--65\% (without the WTA requirement) for $\pTHF <3\gev$ to more than 95\% for $\pTHF > 6\gev$.

Bin migration effects, which originate from tracking inefficiency, material interactions, and finite momentum and energy resolution in the \lhcb detector, are then corrected using a 6D response matrix (RM) that describes the mapping of \pTHF, \pTjet, and \mjet/\pTjet before and after detector effects.
A 3D unfolding is performed using the iterative Bayesian unfolding algorithm~\cite{d1995multidimensional}, as implemented in \textsc{RooUnfold}~\cite{adye2011_RooUnfold, *RooUnfold}. Three iterations are used for all observable configurations, which provides good convergence. Unfolding begins with a prior distribution from simulation before iteratively updating the distribution using Bayes' theorem with the calculated RM and measured data. The bin migration is dominated by a strong diagonal mapping in the RM coupled with a slight smearing along the \pTjet and \mjet/\pTjet axes, with negligible smearing in \pTHF.

The \mjet/\pTjet distribution is then corrected for
jets which are not reconstructed or selected in data due to kinematic or topological selections, or detector inefficiencies. The total \bquark-jet reconstruction efficiency is calculated using simulation and adjusted with data-driven methods. Specifically, the efficiency of reconstructing the \Bpm decay products is calculated using the tag-and-probe method on $\jpsi\to\mup\mun$ decays~\cite{LHCB-DP-2013-002}. The efficiency of correctly identifying these particles is calculated using standard calibration samples~\cite{LHCb-PUB-2016-021}. The trigger efficiency is calculated using \jpsi\to\mup\mun decays selected by several hardware triggers~\cite{LHCb-PUB-2014-039}.
The total efficiency rises from $5\%$ for $\pTHF < 4\gev$ to around 20\% for $\pTHF > 20\gev$.

%%%%%%%%%%%%%%%%%%%%%%%%%%%%%%%%%%%%%%%%%%%%%%%%%%%%%%%%%%%%%%%%%%%
\section{Systematic uncertainties}

To probe the dependence of \mjet/\pTjet on underlying systematic uncertainties, several procedural variations are performed with the full analysis repeated. The ratio between these varied results and the baseline one is used to assign a relative systematic uncertainty.

Tighter selection requirements on the \jpsi and \Bpm candidates are used to test the robustness of the reconstruction and selection efficiency. The signal and background extraction is varied by replacing the DSCB functions with a Student's $t$-distribution. The PB sideband regions are varied by adjusting the lower and upper edges of both sidebands, and the uncertainties arising from these variations are averaged.
Uncertainties on the tracking and PID efficiencies for the \Bpm candidate decay products are propagated from statistical uncertainties on the calibration samples and the uncertainty on the material budget for the \Kpm meson, which is estimated at $\pm1.4$\%~\cite{LHCB-DP-2013-002}.
The uncertainty on the trigger efficiency is estimated by applying the data-driven method in simulation and comparing the result to the true simulated efficiency. Any deviation of this ratio from unity is propagated as a shift in the trigger efficiency.

The uncertainty on the jet energy scale and resolution is determined by studying jets recoiling off a \Z boson, reconstructed via its decay to \mup\mun. The relative transverse momentum fraction $\pTjet/p_{\mathrm{T},\Z}$ is calculated in both data and detector-level simulation, and the simulated distribution is adjusted to fit the data distribution as a function of \pTjet. This yields a roughly $-3\%$ shift on the jet energy scale and $+12\%$ smearing in the jet energy resolution, which are propagated to the detector-level simulation used to calculate detector corrections. Since \mjet explicitly depends on $E_\mathrm{jet}$ and $p_\mathrm{jet}$, this uncertainty tends to be dominant.

Finally, the uncertainty on detector corrections is estimated. The number of iterations through the unfolding procedure is varied by $\pm1$ iteration.
The dependence on the prior distribution is probed by weighting the RM by the ratio of sideband-subtracted data to detector-level simulation.
A series of closure tests is also performed, where detector corrections are applied on a simulated data spectrum smeared by statistical uncertainties in data. Any statistically significant deviations from unity in the closure tests are taken as a bin-by-bin systematic uncertainty. The total unfolding uncertainty is calculated by assuming these underlying variations to be correlated and calculating their standard deviation from zero.

The total systematic uncertainty is then calculated by assuming that each contribution is independent and summing them in quadrature. The systematic uncertainties generally increase with \pTjet and are largest in the sparsely populated tails of the \mjet/\pTjet distribution, where small changes in the peak position cause the largest effect. Systematic uncertainties are dominant at low and moderate \pTjet, whereas statistical uncertainties become more significant in the higher \pTjet bins. See the supplementary material for a summary table of the systematic uncertainties.

%%%%%%%%%%%%%%%%%%%%%%%%%%%%%%%%%%%%%%%%%%%%%%%%%%%%%%%%%%%%%%%%%%%
\section{Results}

The \mjet/\pTjet distributions are reported in six \pTjet bins between $10 < \pTjet < 100$\gev. The distributions are normalised per \pTjet bin,
\begin{equation}
\frac{1}{N_\mathrm{jet}} \frac{\mathrm{d}N_\mathrm{jet}}{\mathrm{d}(\mjet/\pTjet)},
\hspace{1em} \mathrm{or\; equivalently} \hspace{1em}
\frac{1}{\sigma_\mathrm{jet}} \frac{\mathrm{d}\sigma_\mathrm{jet}}{\mathrm{d}(\mjet/\pTjet)},
\end{equation}
where $N_\mathrm{jet}$ is the number of jets in a given \pTjet bin, and $\sigma_\mathrm{jet}$ is the corresponding cross-section.

The ungroomed \mjet/\pTjet distributions are shown in Fig.~\ref{fig:mjet_ungr_WTAcomp} for both inclusive and WTA flavour-tagged \bquark jets, with normalisation set to the inclusive results. The peak of the distribution moves leftward with increasing \pTjet due to the reduced contribution of the \bquark-quark mass to the overall jet mass. There is a long tail on the right side of the distributions, corresponding to rarer, higher-virtuality partons which initiate the \bquark jet, and a shorter tail on the left side of the distributions, terminating at the kinematic cutoff of $m_{B^\pm} / \pTjet$ (a single-particle \bquark jet). At high \pTjet, where the effect of gluon splitting is largest, a two-peak structure emerges, which likely originates from the presence of a second (unreconstructed) \bquark-hadron in the jet.

The WTA flavour tag changes the shape of the distributions by reducing contributions in the high-mass tail, and these differences become larger at higher \pTjet. This is expected since the WTA flavour tag tends to reject \bquark jets that originate from gluon splitting, which naturally must have a mass greater than two times the \bquark-quark mass.

The distributions are compared to \pythia 8 predictions~\cite{Sjostrand:2007gs}, where significant deviations are seen. \pythia 8 predicts a broader \mjet/\pTjet distribution, resulting in significantly more emissions at large \mjet/\pTjet than observed in data. It also peaks at lower values, signifying fewer high-virtuality partons than observed in data. Differences are enhanced for inclusive distributions; since WTA tagging predominantly removes \bquark-jets that arise from gluon splitting~\cite{Les_Houches_studies}, this suggests that in-shower contributions to \bquark-jet production are less constrained than direct (LO) \bquark-jet production.

\afterpage{%
\begin{figure}[tb]
    \centering
    \includegraphics[page=1,width=0.49\linewidth]{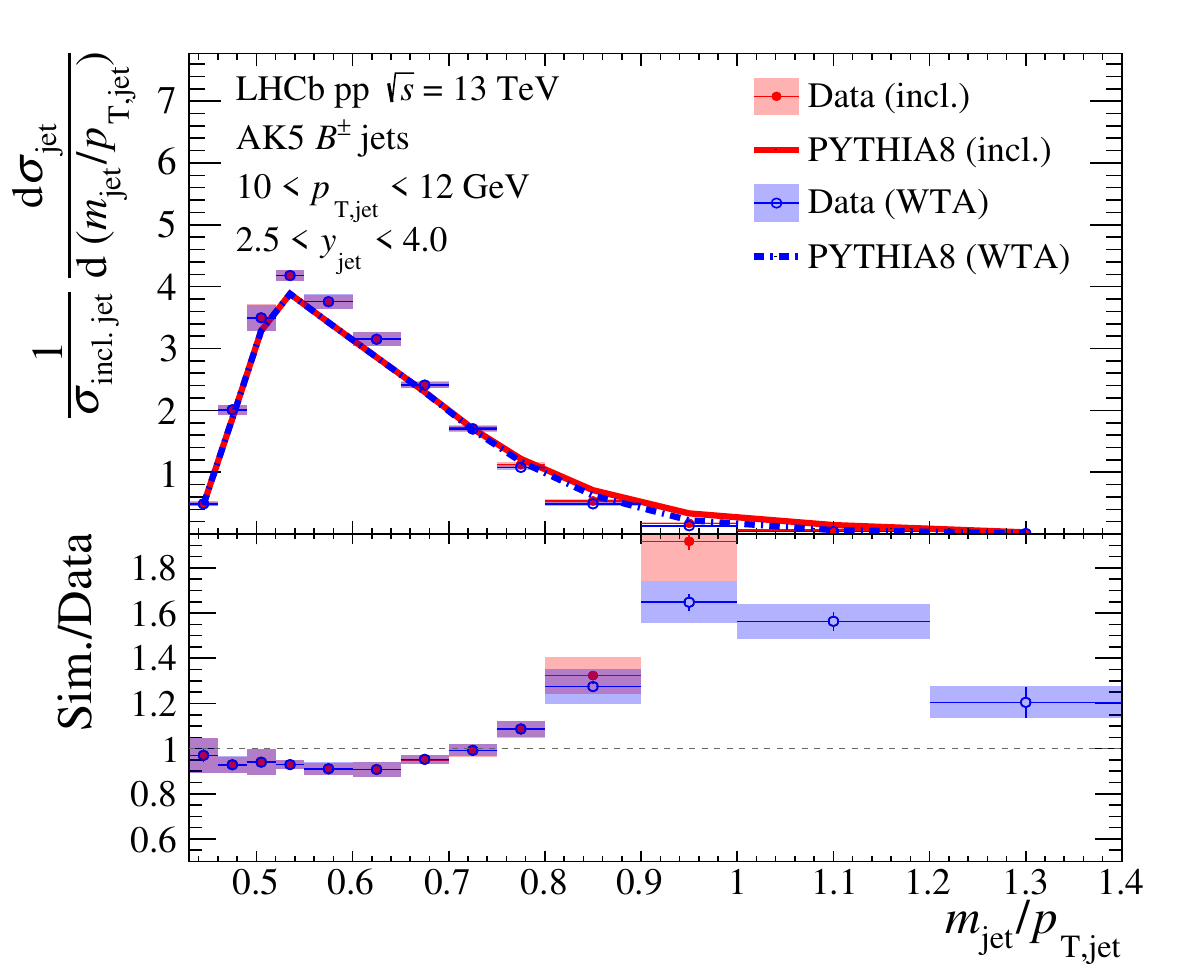}
    \includegraphics[page=1,width=0.49\linewidth]{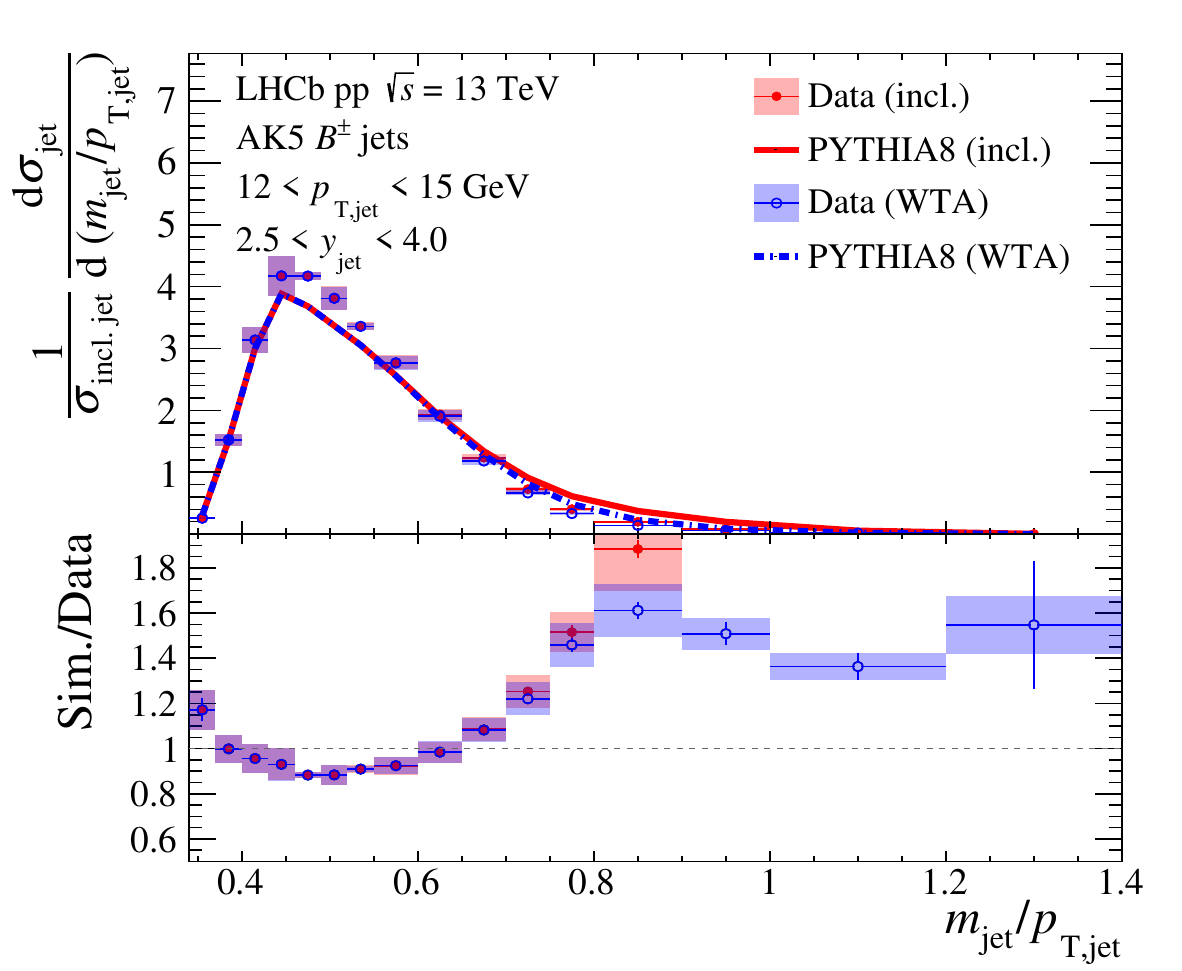}
    \includegraphics[page=1,width=0.49\linewidth]{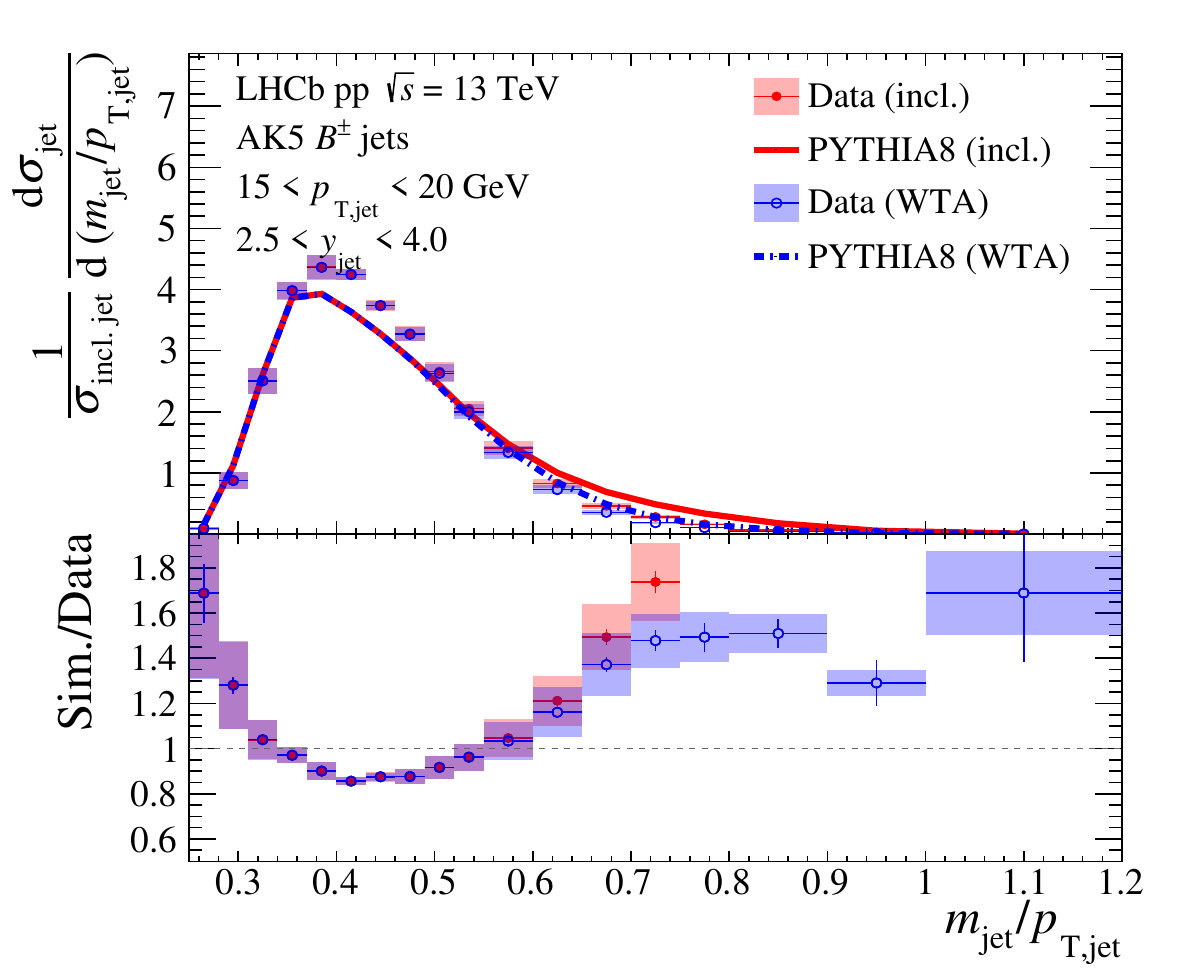}
    \includegraphics[page=1,width=0.49\linewidth]{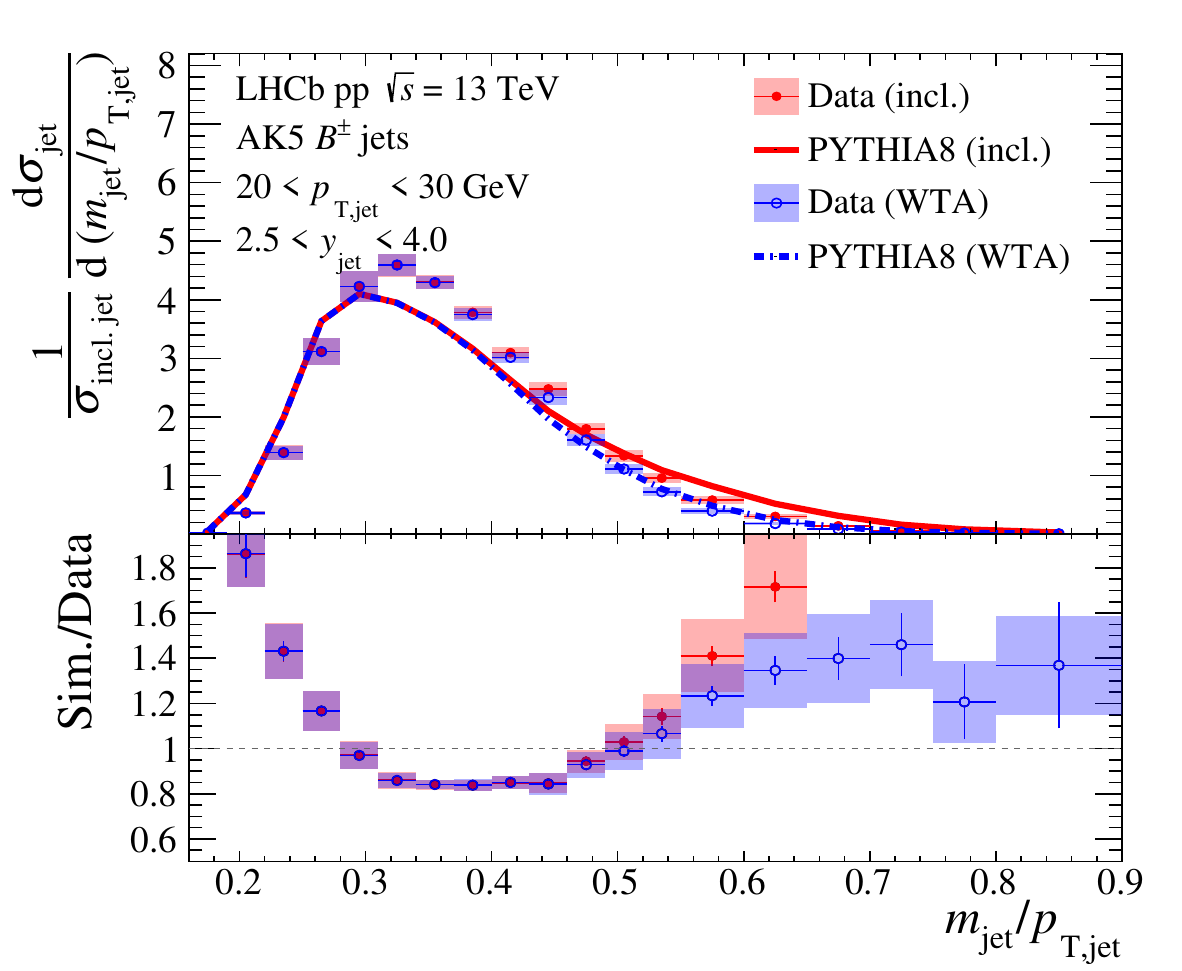}
    \includegraphics[page=1,width=0.49\linewidth]{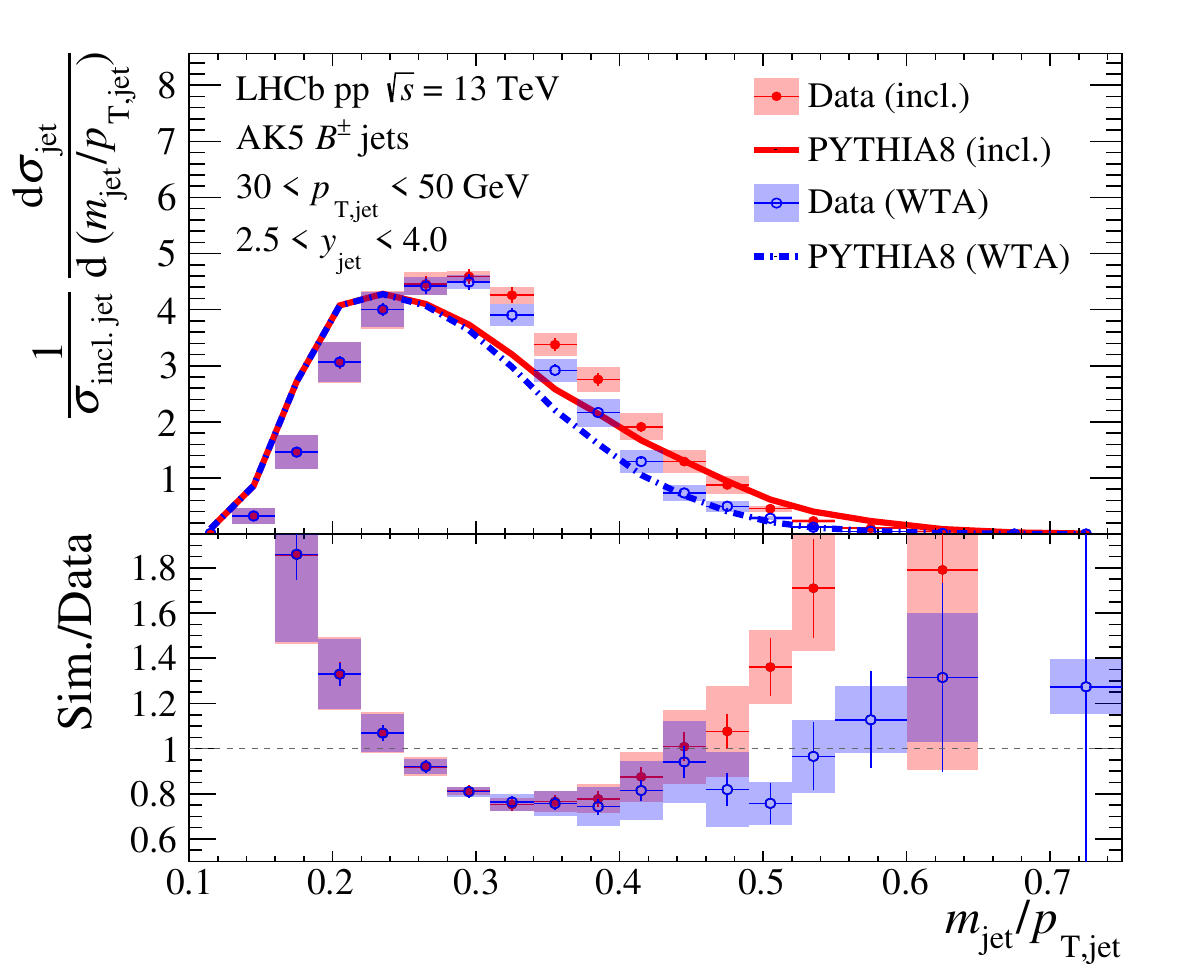}
    \includegraphics[page=1,width=0.49\linewidth]{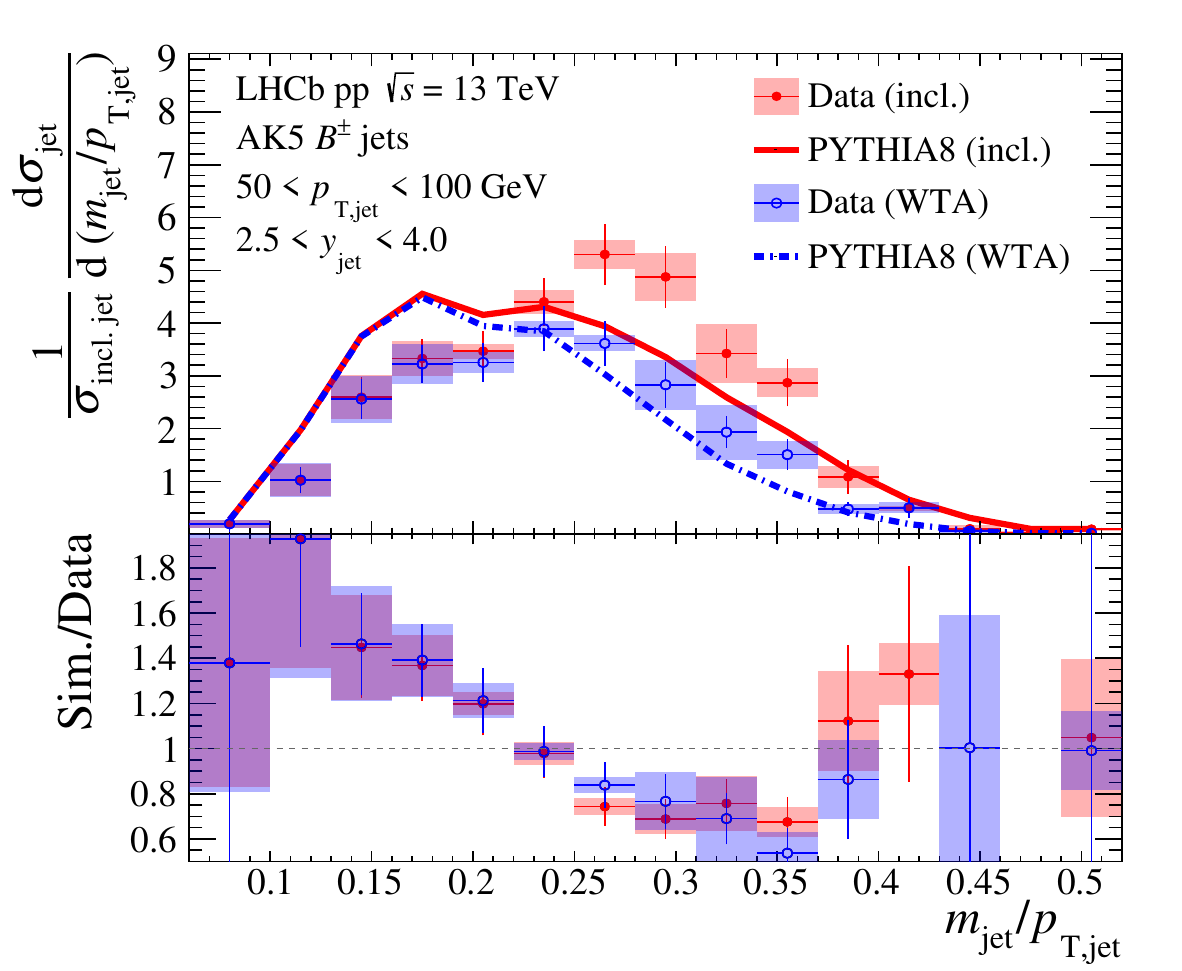}
    \caption{Ungroomed $m_\mathrm{jet} / p_\mathrm{T,jet}$ distributions for $R=0.5$ anti-$k_\mathrm{T}$ jets (AK5) containing a $B^\pm$ meson with (WTA) and without (incl.)~WTA flavour tagging. Results are shown for several bins of $p_\mathrm{T,jet}$, ranging from 10 to 100\gev, and are compared to predictions from \pythia 8.186 (\lhcb tune). The shaded error bands correspond to total systematic uncertainties, while the solid lines correspond to statistical uncertainties. Some points are not displayed in the ratio panel because they are outside the plotted range. Note that the $x$-axis range changes for each $p_\mathrm{T,jet}$ bin, due to quark-mass effects.}
    \label{fig:mjet_ungr_WTAcomp}
\end{figure}

\begin{figure}[tb]
    \centering
    \includegraphics[page=1,width=0.49\linewidth]{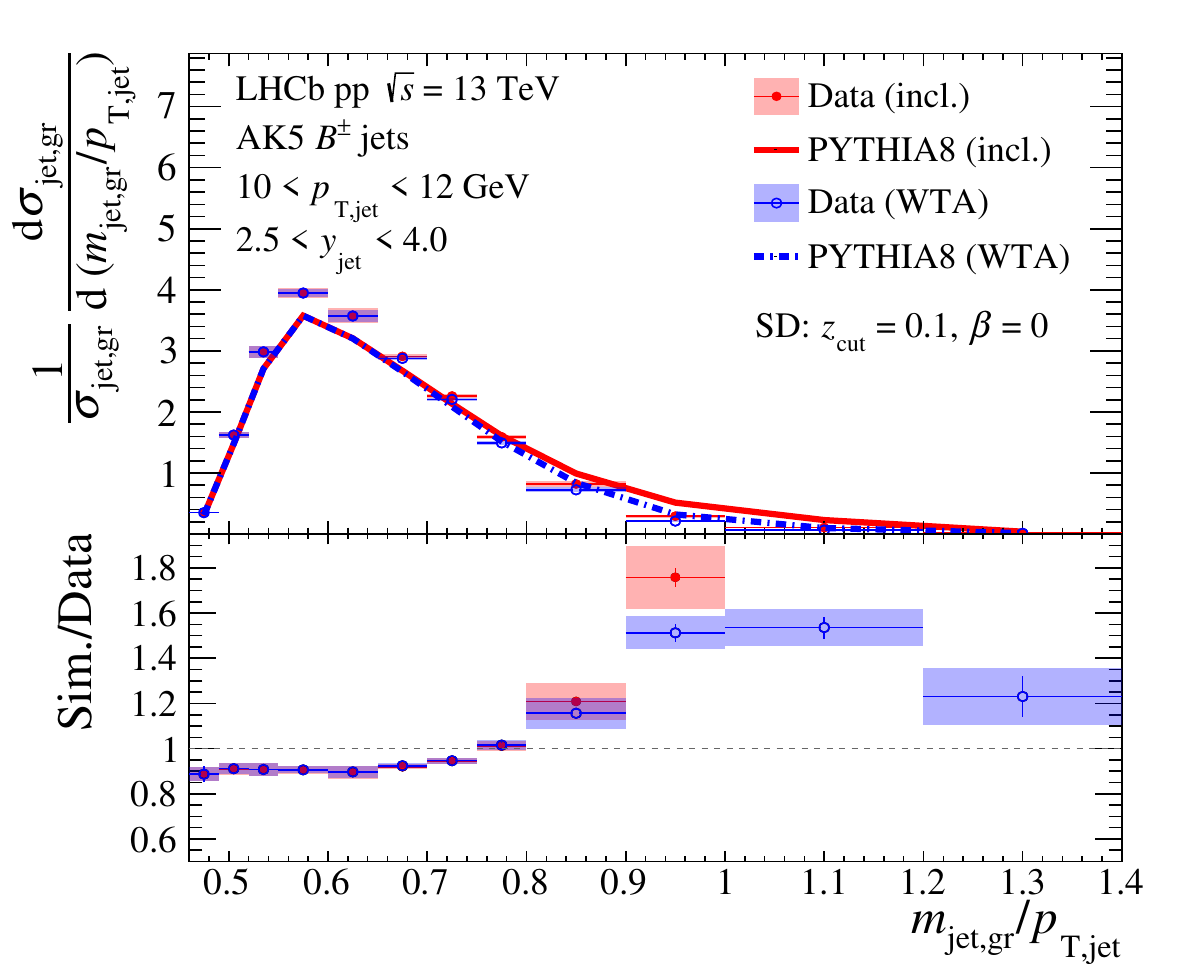}
    \includegraphics[page=1,width=0.49\linewidth]{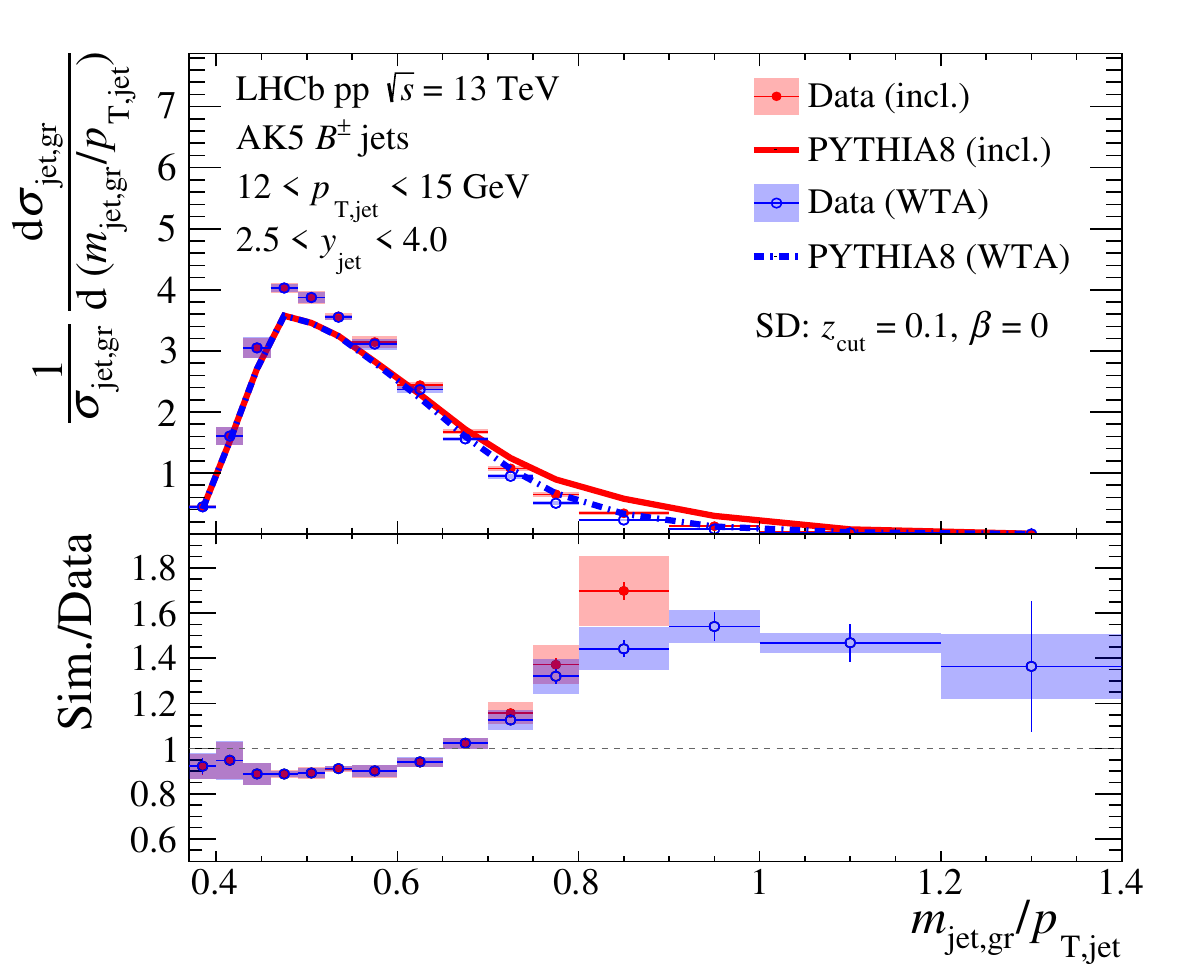}
    \includegraphics[page=1,width=0.49\linewidth]{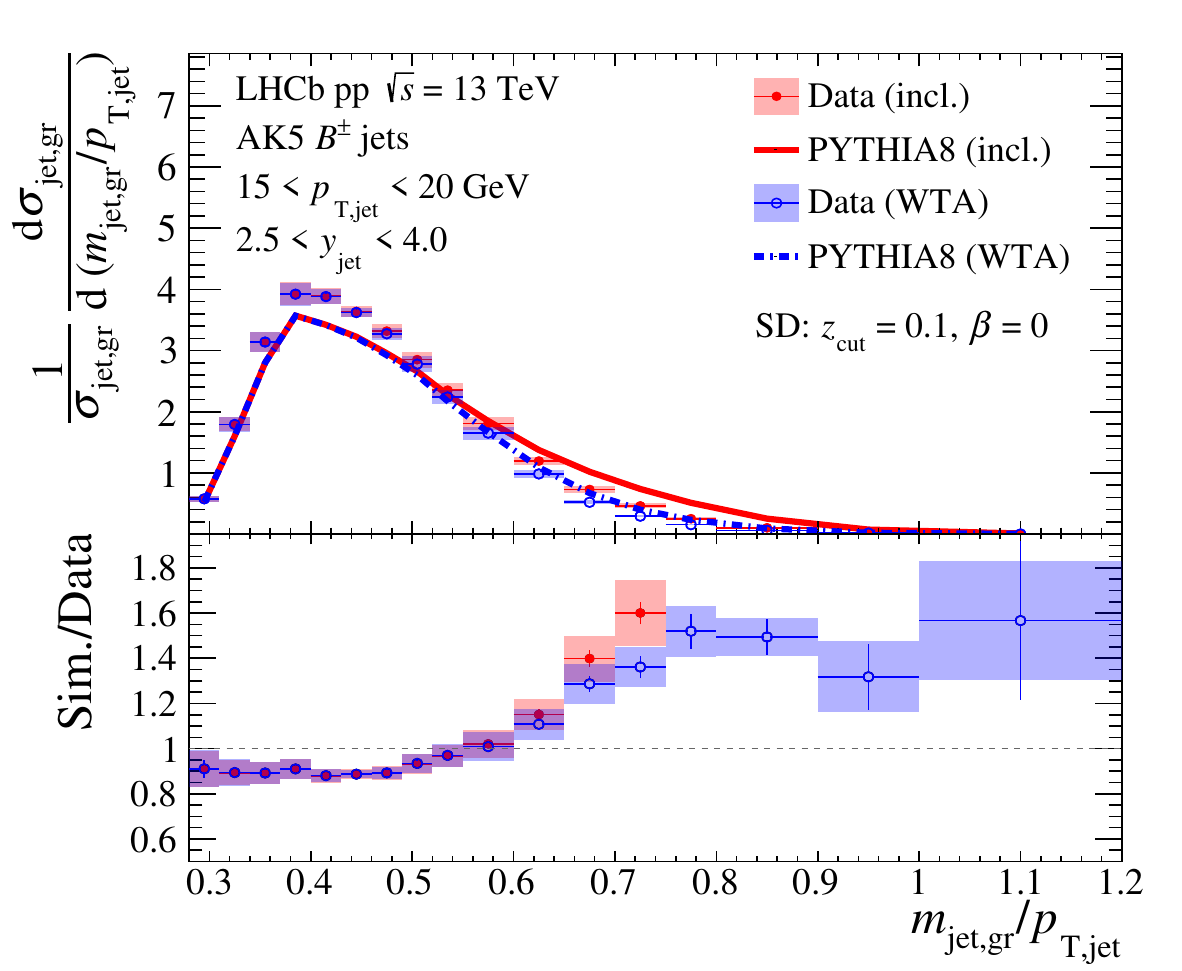}
    \includegraphics[page=1,width=0.49\linewidth]{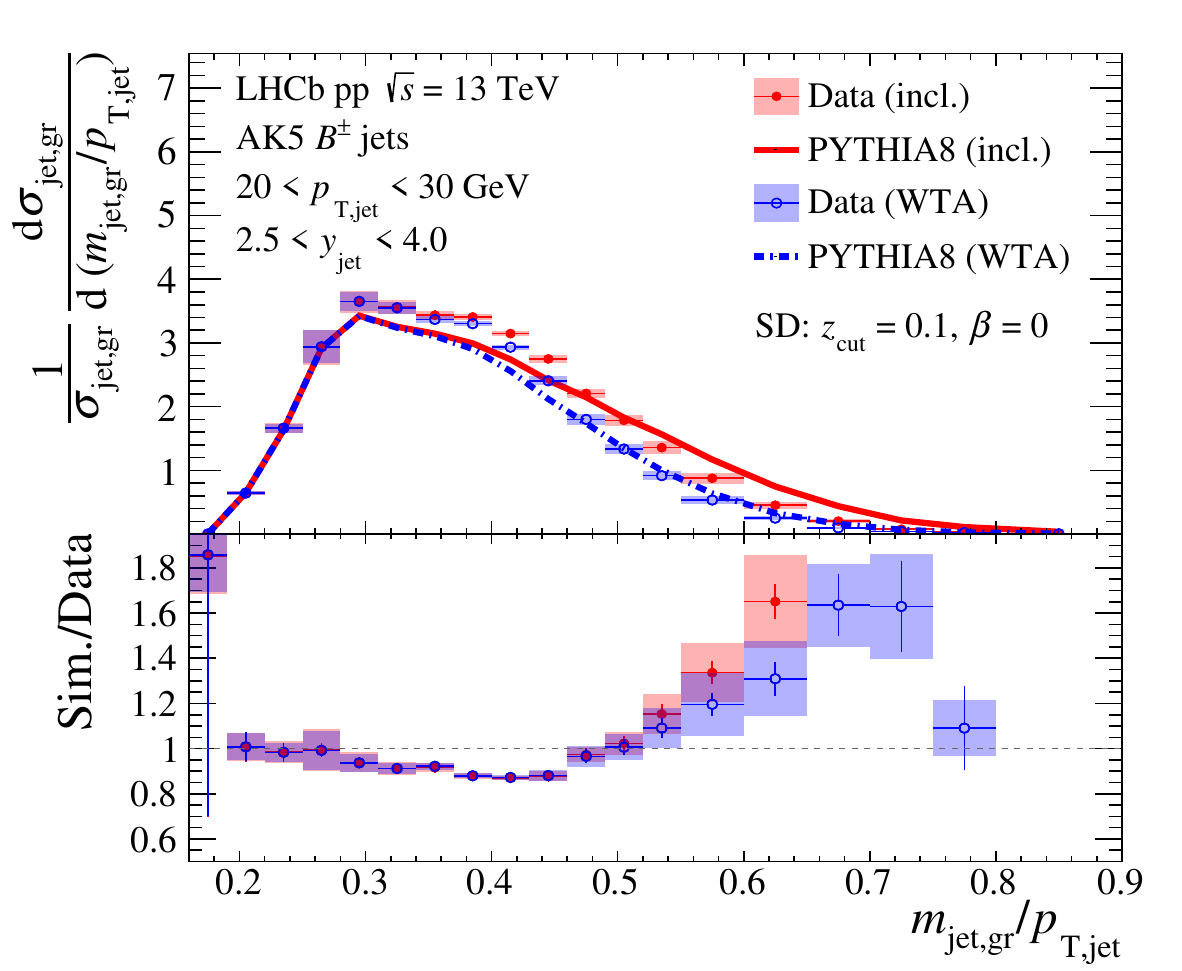}
    \includegraphics[page=1,width=0.49\linewidth]{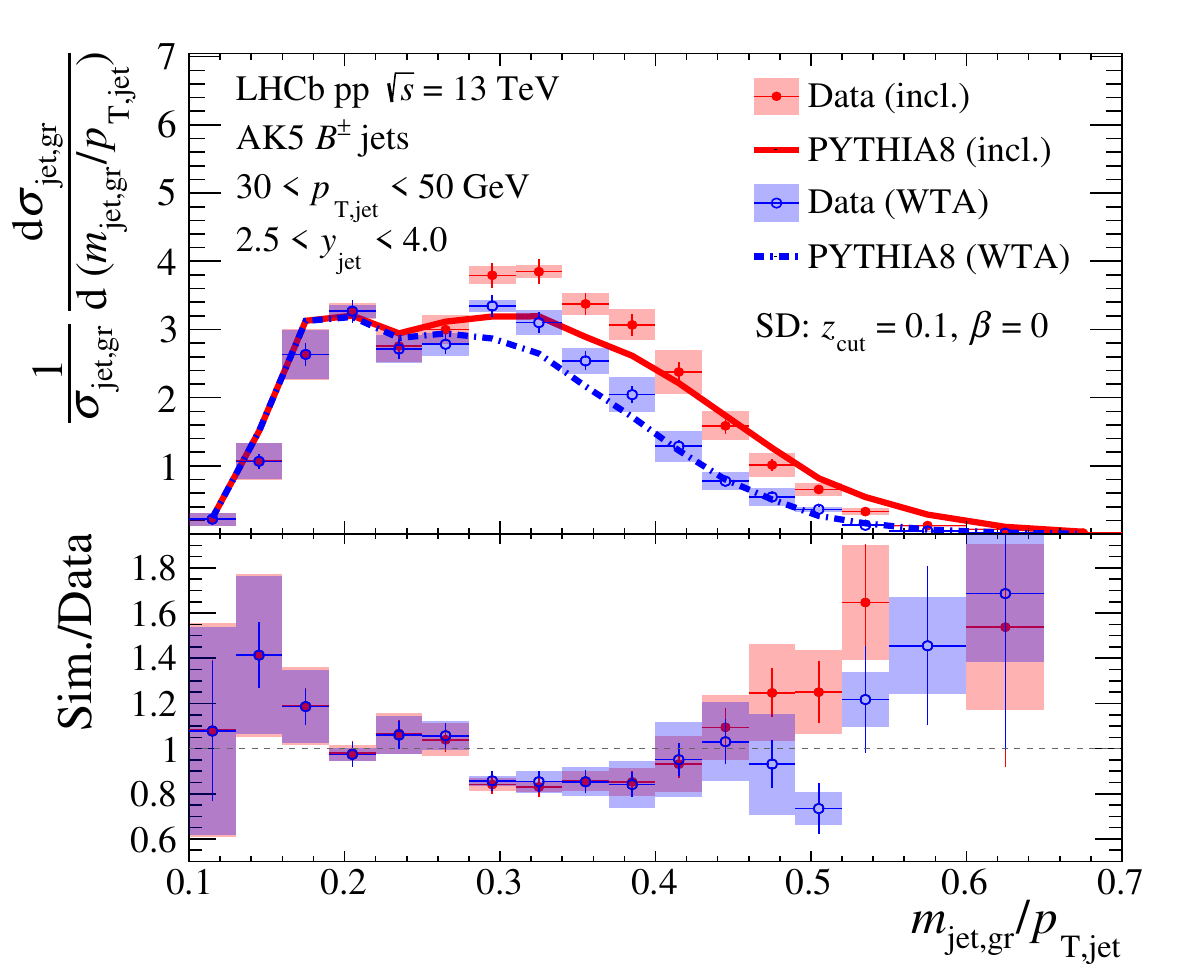}
    \includegraphics[page=1,width=0.49\linewidth]{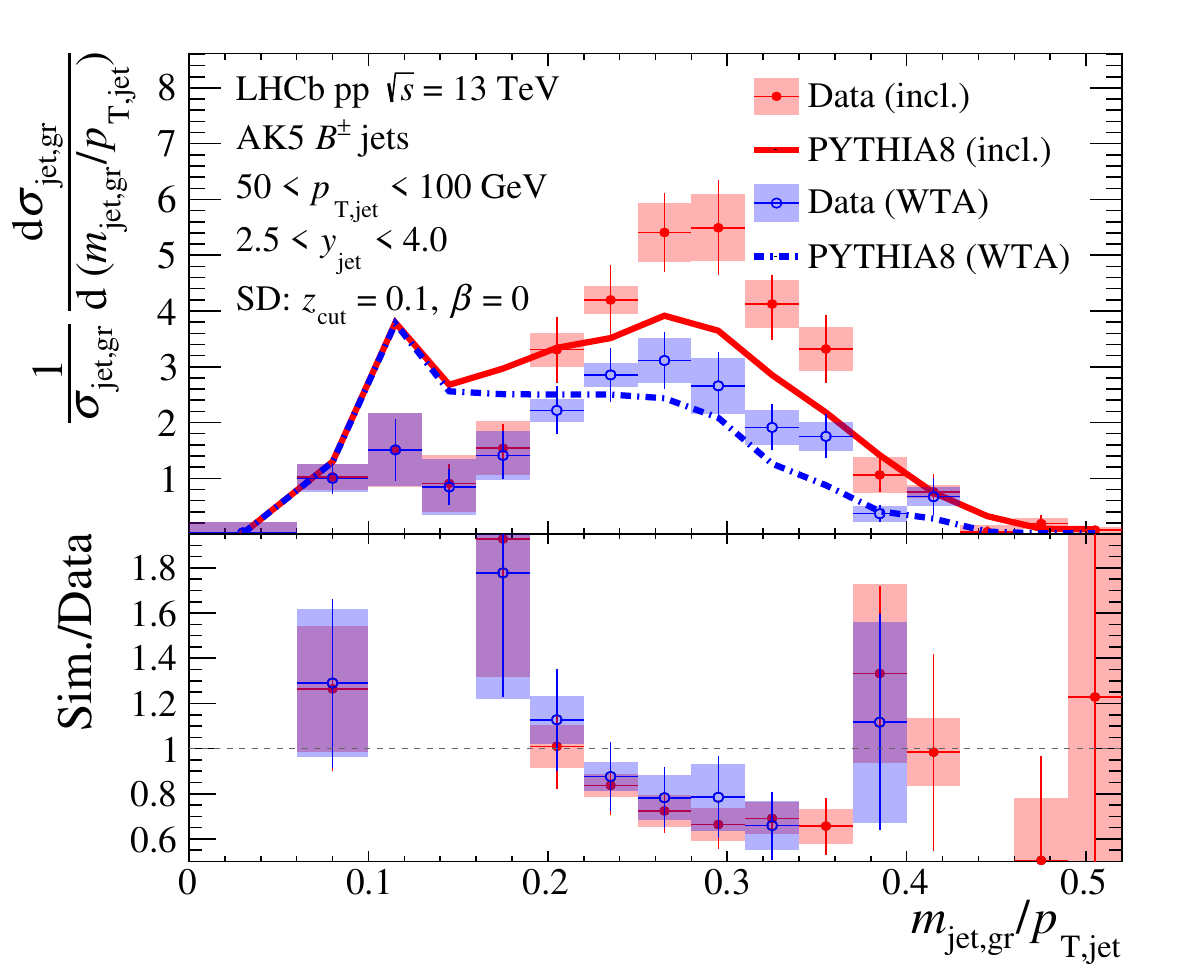}
    \caption{Soft drop groomed $m_\mathrm{jet,gr} / p_\mathrm{T,jet}$ distributions for $R=0.5$ anti-$k_\mathrm{T}$ (AK5) jets containing a $B^\pm$ meson with (WTA) and without (incl.)~WTA flavour tagging. Results are shown for several bins of $p_\mathrm{T,jet}$, ranging from 10 to 100\gev, and are compared to predictions from \pythia 8.186 (\lhcb tune). The shaded error bands correspond to total systematic uncertainties, while the solid lines correspond to statistical uncertainties. Some points are not displayed in the ratio panel because they are outside the plotted range. Note that the $x$-axis range changes for each $p_\mathrm{T,jet}$ bin, due to quark mass effects.}
    \label{fig:mjet_gr_WTAcomp}
\end{figure}
\clearpage

\begin{figure}
    \centering
    \includegraphics[page=1,width=0.49\linewidth]{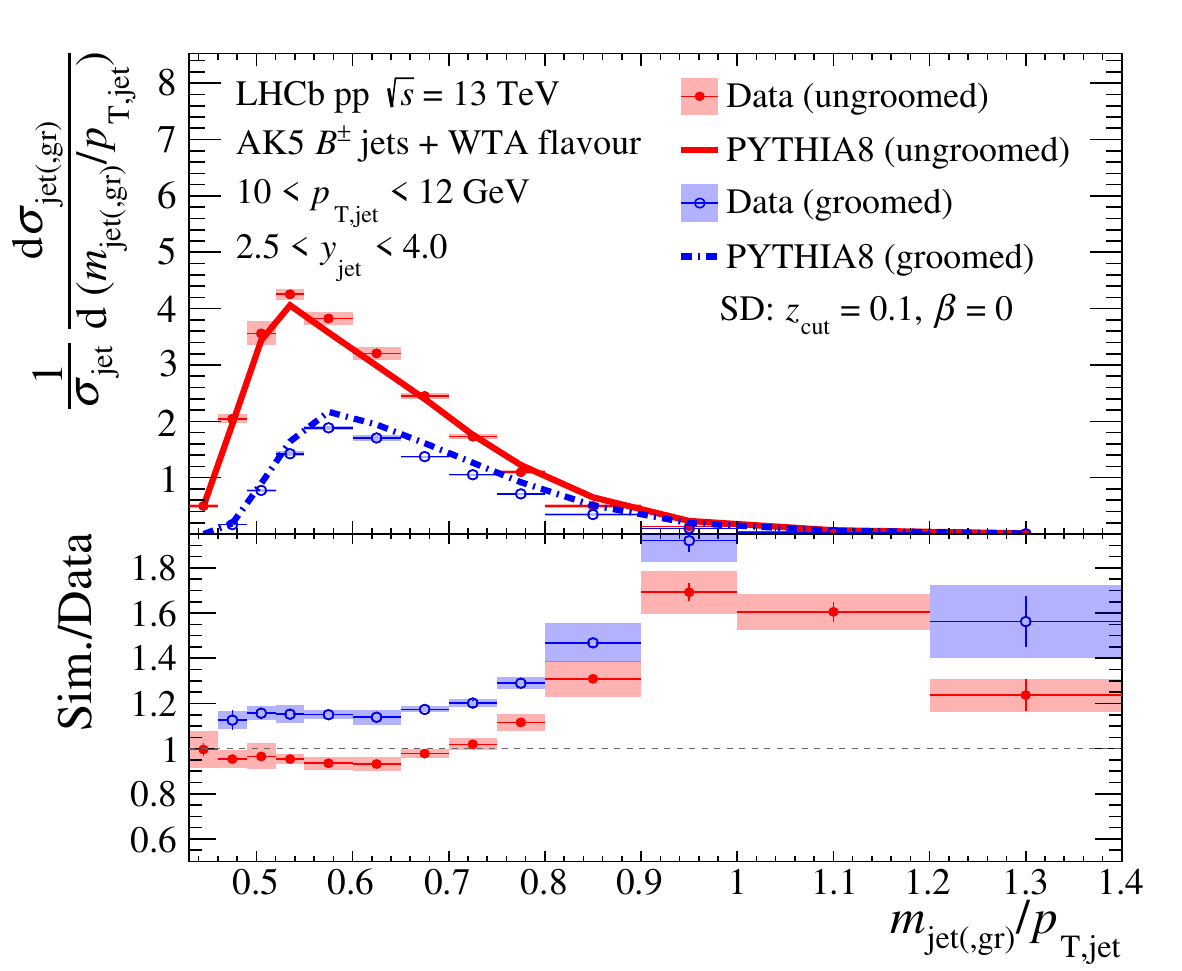}
    \includegraphics[page=1,width=0.49\linewidth]{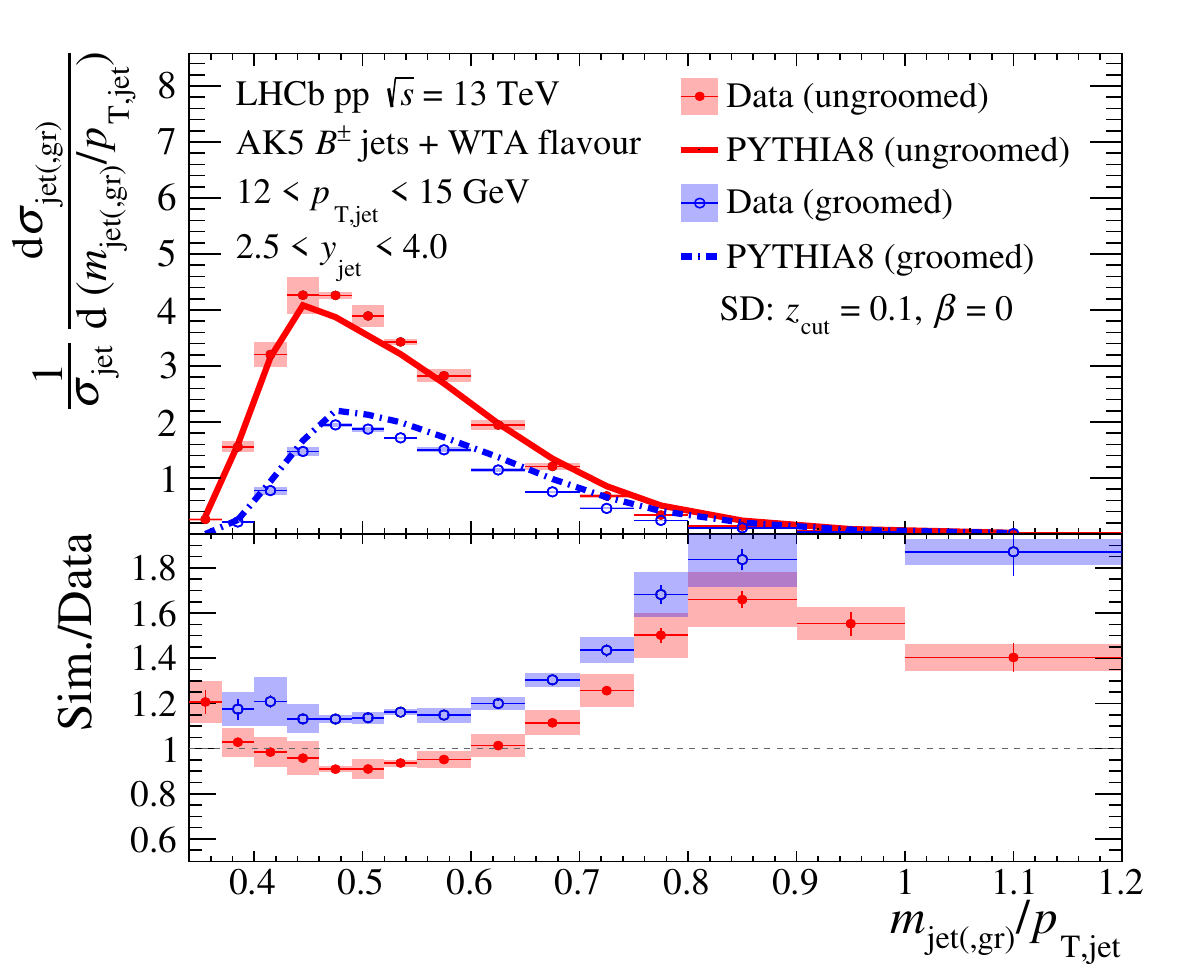}
    \includegraphics[page=1,width=0.49\linewidth]{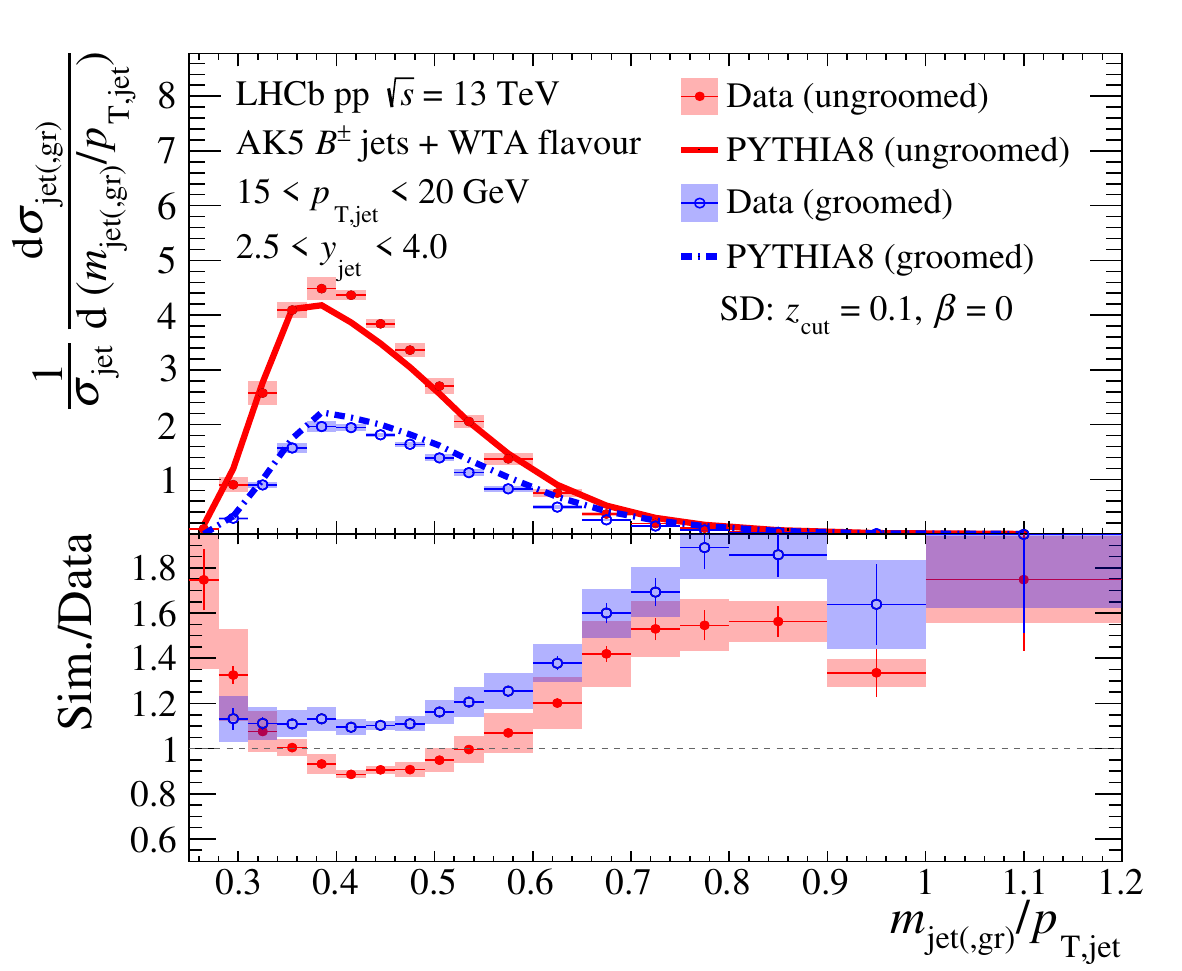}
    \includegraphics[page=1,width=0.49\linewidth]{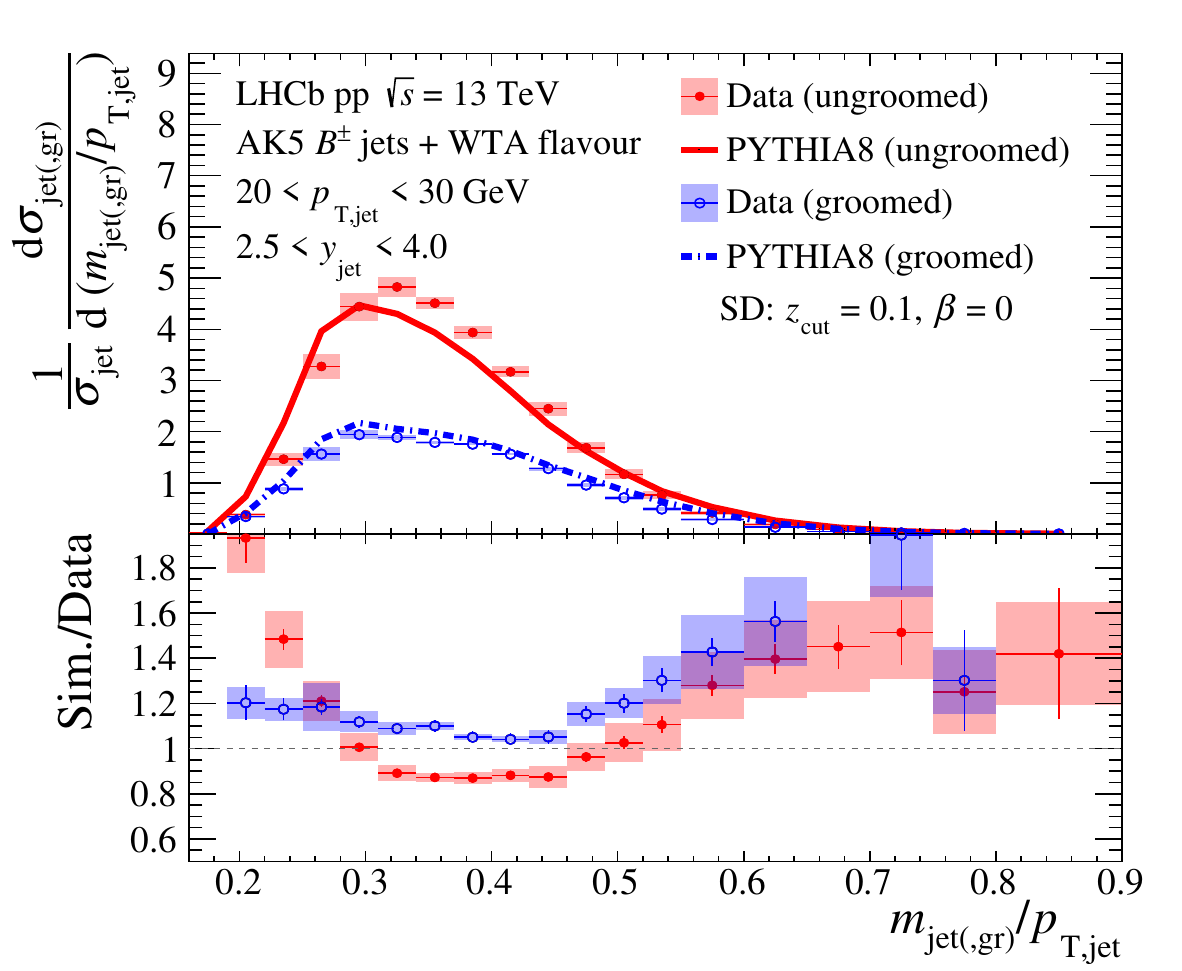}
    \includegraphics[page=1,width=0.49\linewidth]{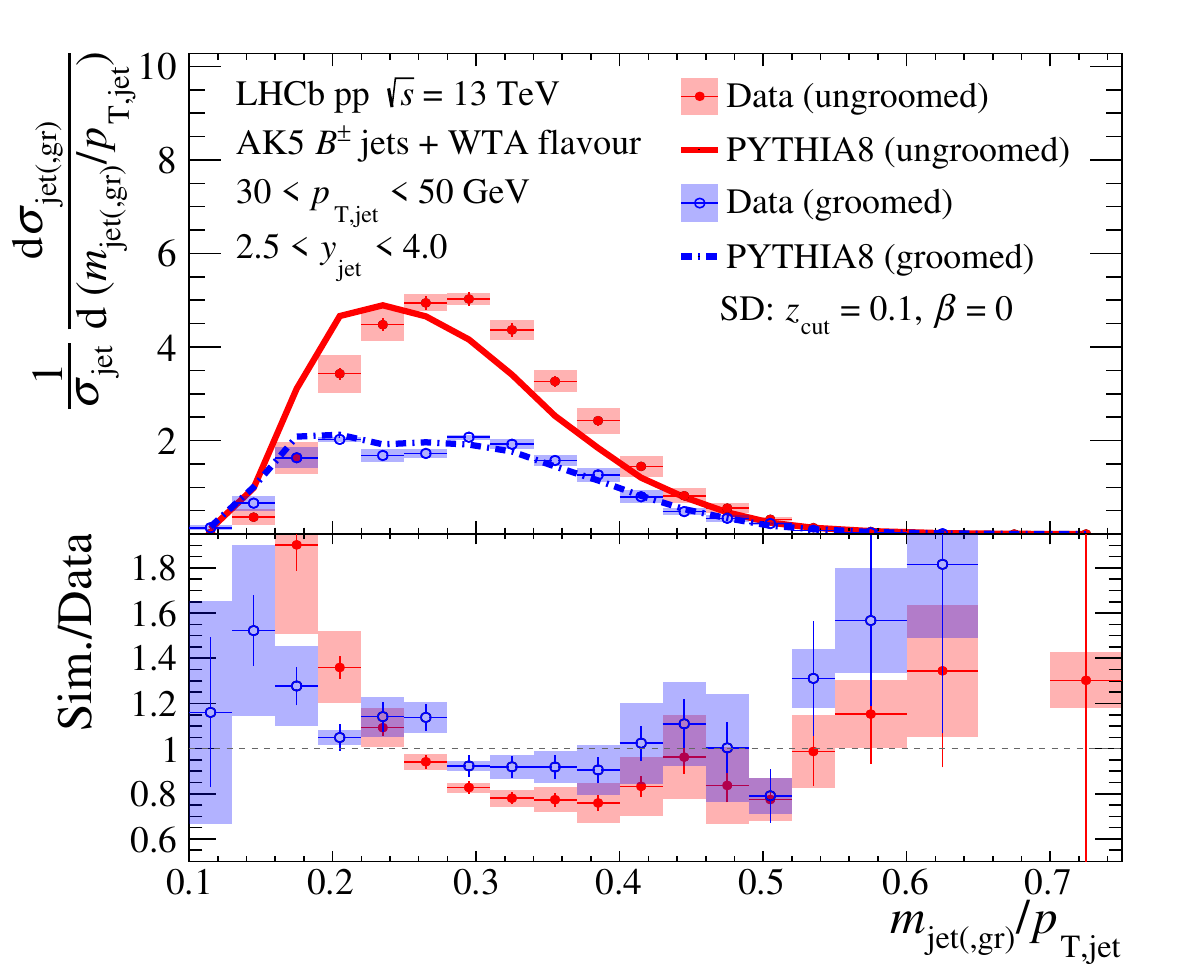}
    \includegraphics[page=1,width=0.49\linewidth]{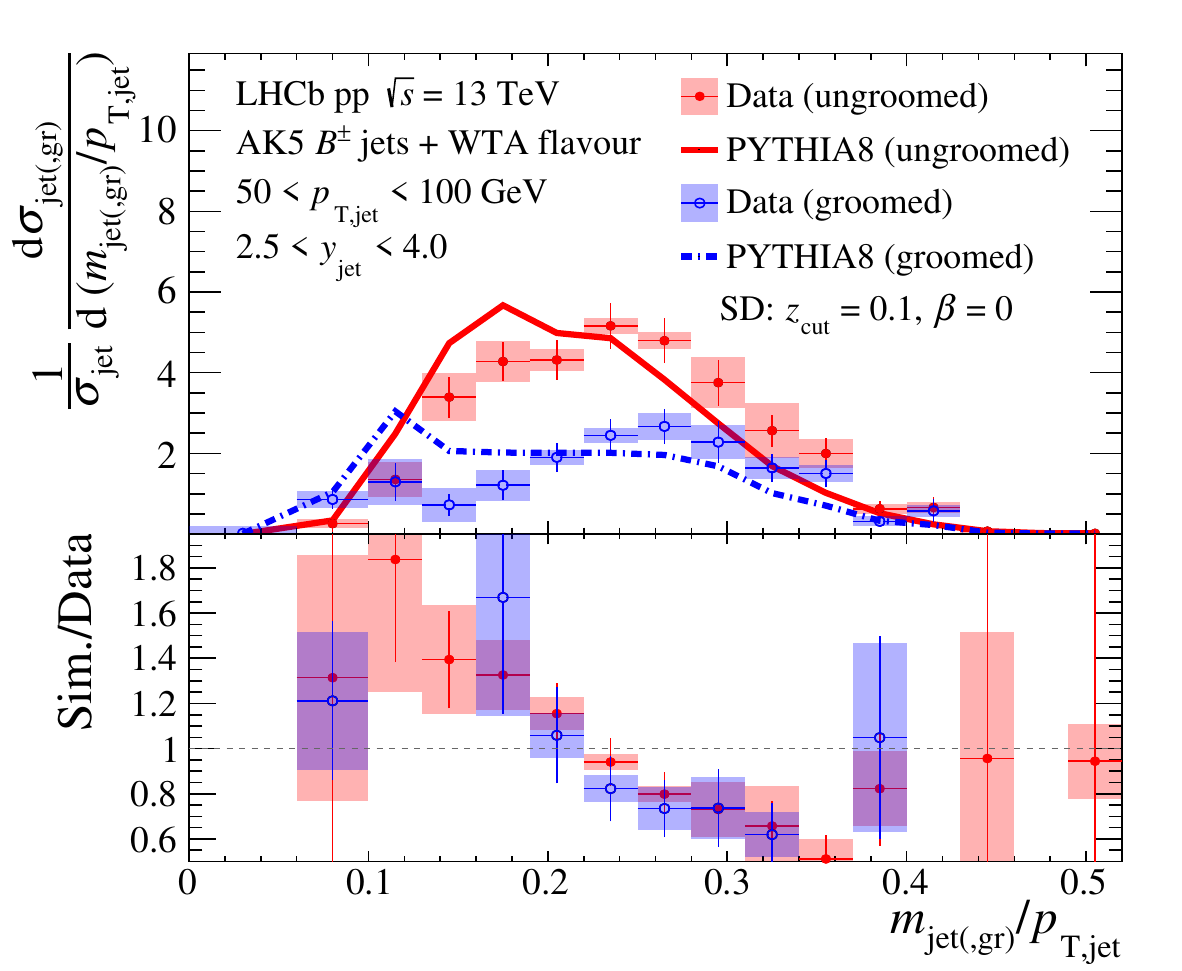}
    \caption{Comparison of ungroomed and soft drop groomed $m_\mathrm{jet(,gr)} / p_\mathrm{T,jet}$ distributions for $R=0.5$ anti-$k_\mathrm{T}$ (AK5) jets containing a $B^\pm$ meson and tagged with WTA flavour tagging. Results are shown for several bins of $p_\mathrm{T,jet}$, ranging from 10 to 100\gev, and are compared to predictions from \pythia 8.186 (\lhcb tune). The shaded error bands correspond to total systematic uncertainties, while the solid lines correspond to statistical uncertainties. Some points are not displayed in the ratio panel because they are outside the plotted range. Note that the $x$-axis range changes for each $p_\mathrm{T,jet}$ bin, due to quark mass effects.}
    \label{fig:mjet_grcomp_WTA}
\end{figure}
\clearpage

\begin{figure}
    \centering
    \includegraphics[page=1,width=0.49\linewidth]{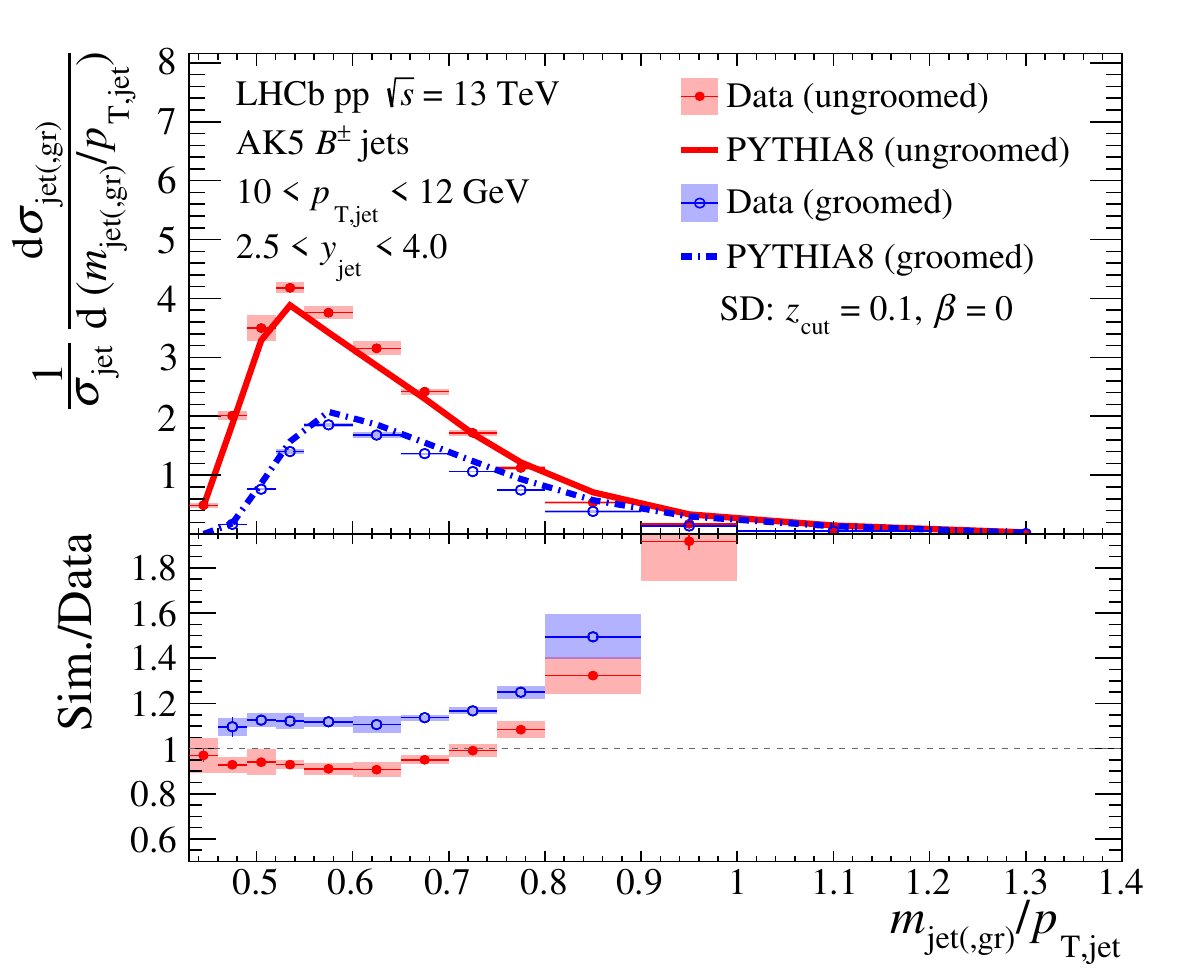}
    \includegraphics[page=1,width=0.49\linewidth]{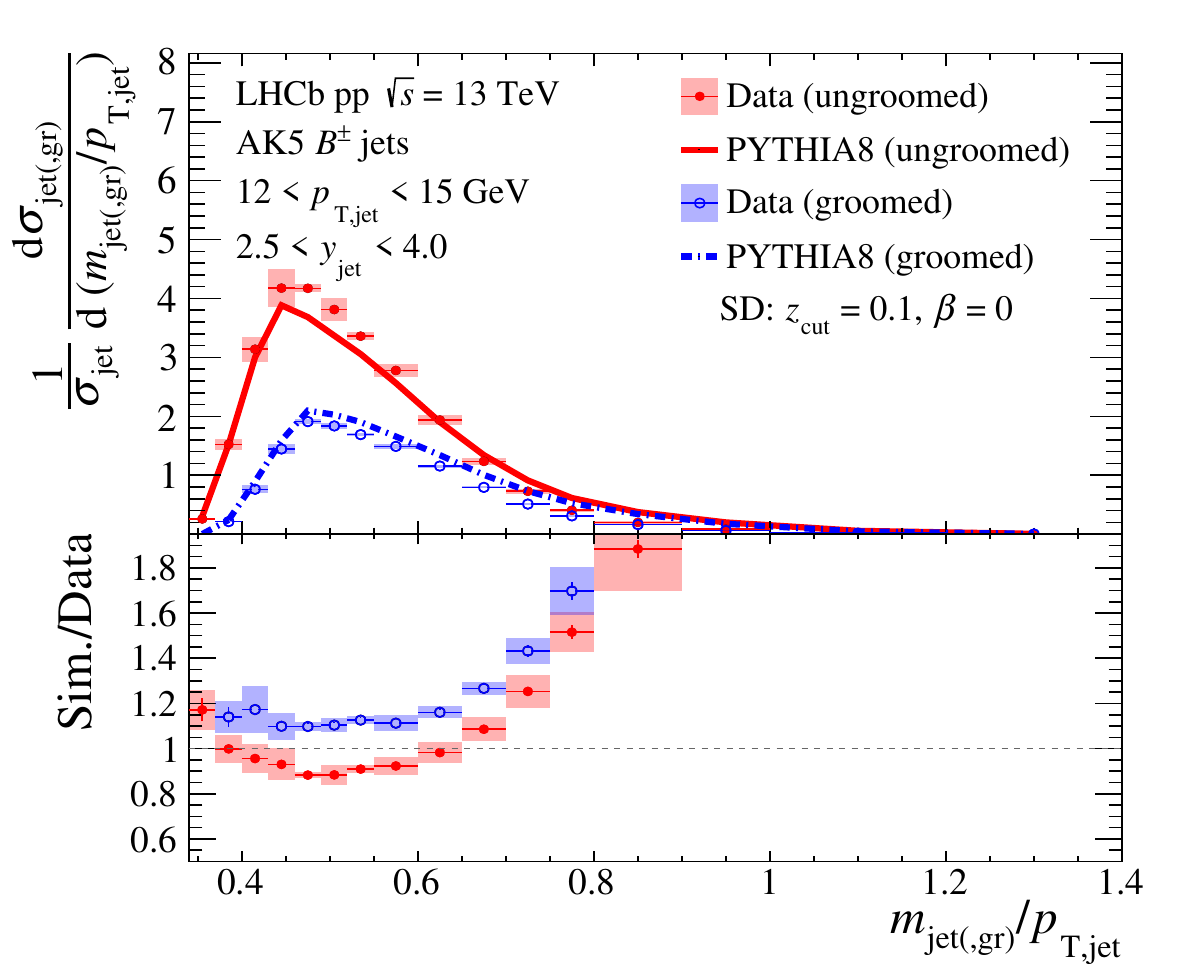}
    \includegraphics[page=1,width=0.49\linewidth]{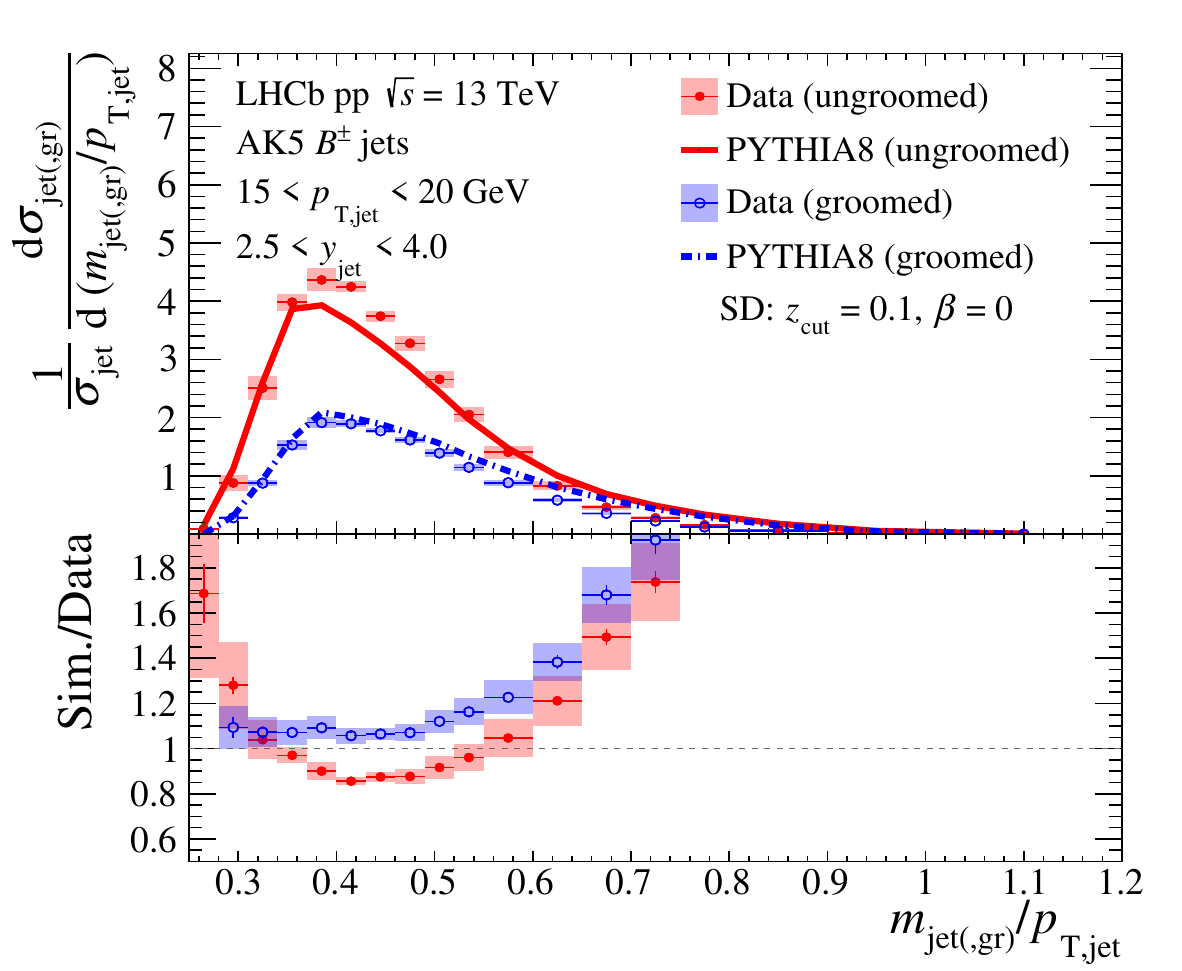}
    \includegraphics[page=1,width=0.49\linewidth]{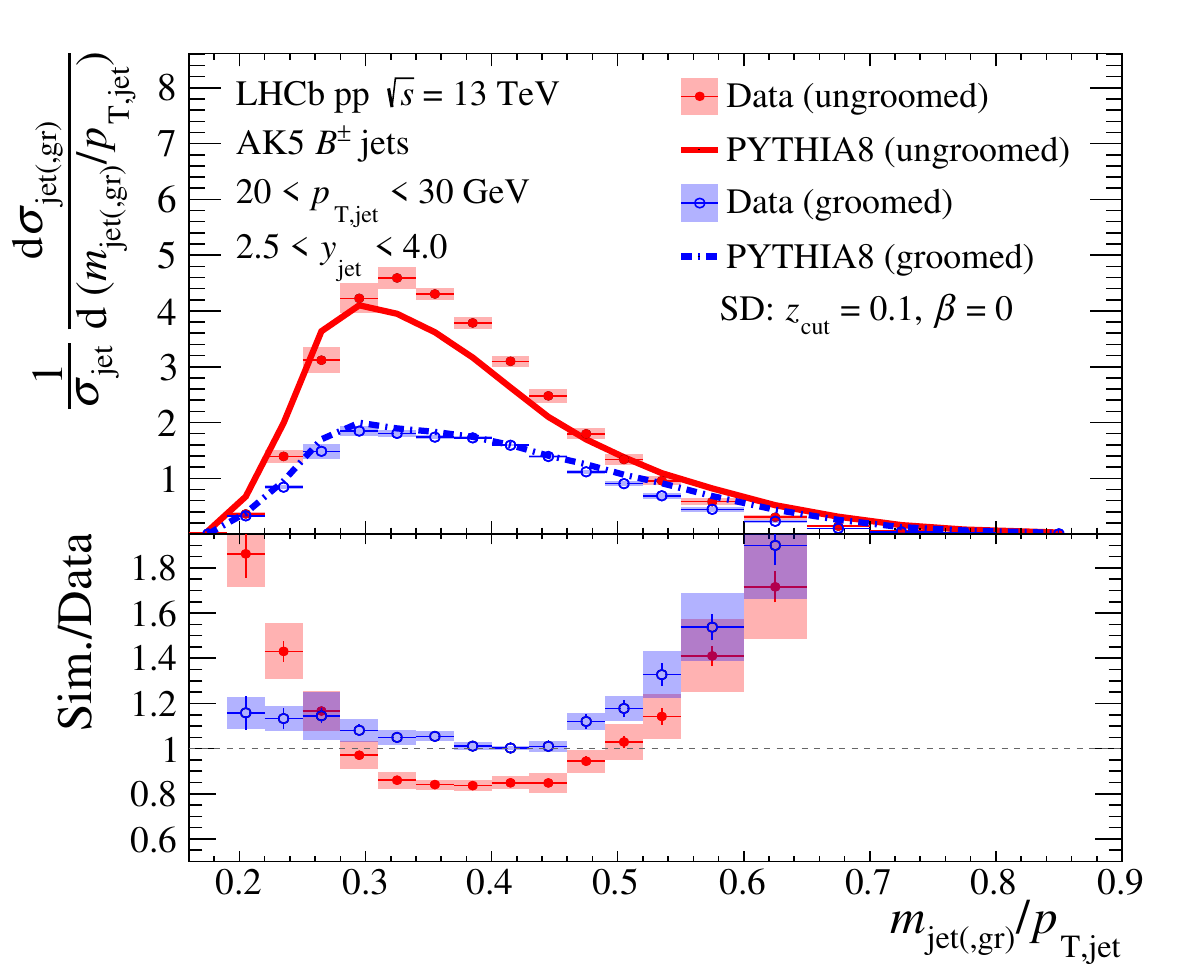}
    \includegraphics[page=1,width=0.49\linewidth]{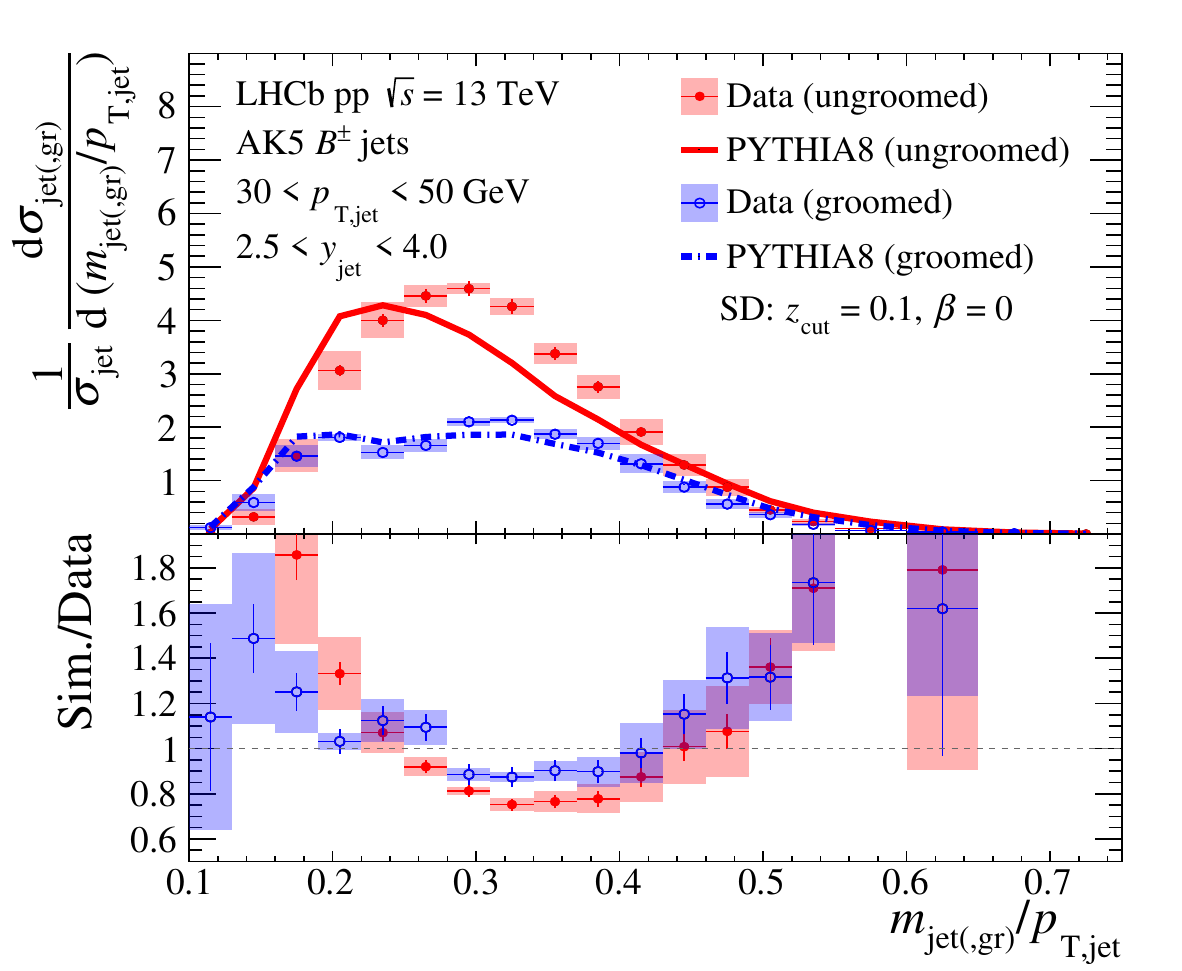}
    \includegraphics[page=1,width=0.49\linewidth]{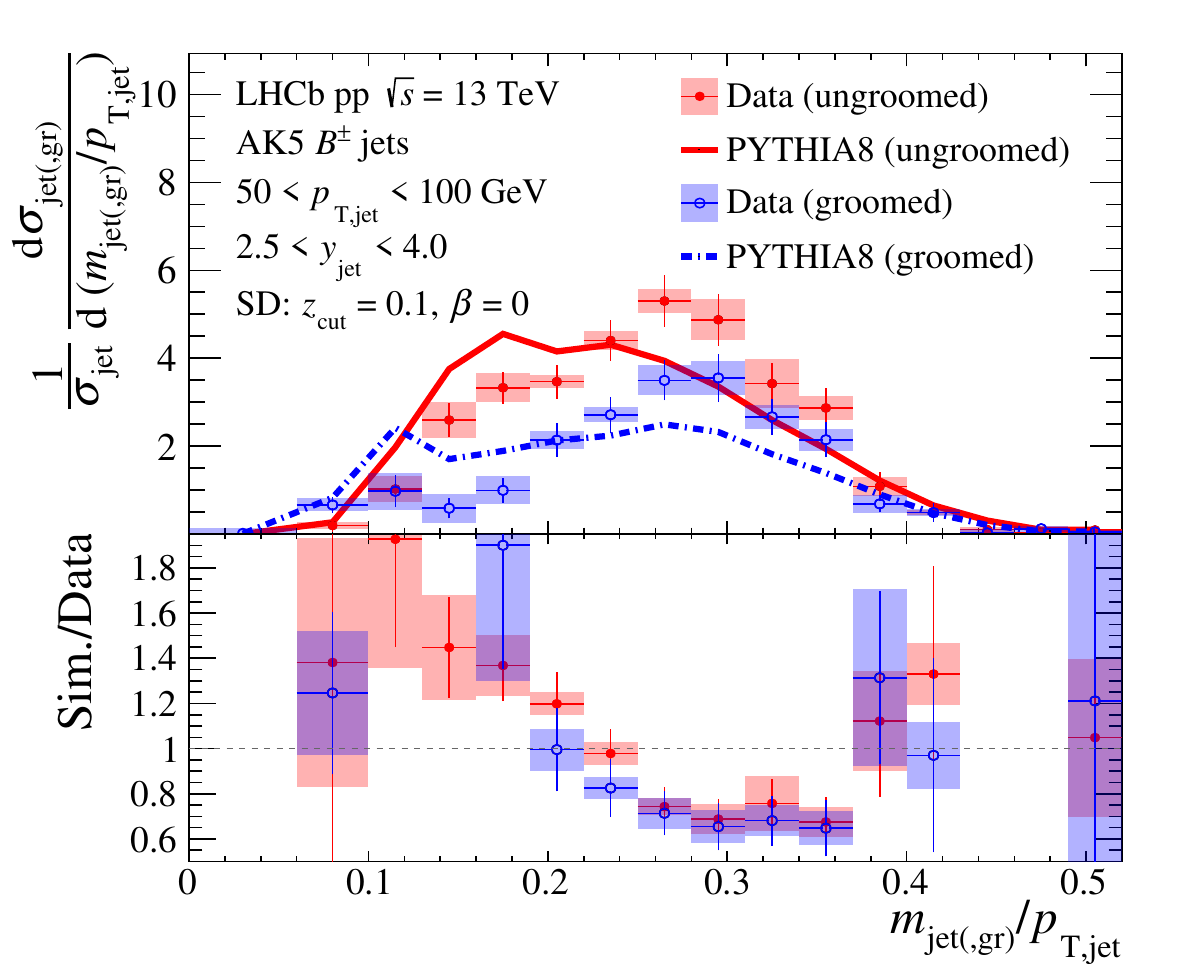}
    \caption{Comparison of ungroomed and soft drop groomed $m_\mathrm{jet(,gr)} / p_\mathrm{T,jet}$ distributions for $R=0.5$ anti-$k_\mathrm{T}$ (AK5) jets containing a $B^\pm$ meson without using WTA flavour tagging. Results are shown for several bins of $p_\mathrm{T,jet}$, ranging from 10 to 100\gev, and are compared to predictions from \pythia 8.186 (\lhcb tune). The shaded error bands correspond to total systematic uncertainties, while the solid lines correspond to statistical uncertainties. Some points are not displayed in the ratio panel because they are outside the plotted range. Note that the $x$-axis range changes for each $p_\mathrm{T,jet}$ bin, due to quark mass effects.}
    \label{fig:mjet_grcomp_noWTA}
\end{figure}
\clearpage
}

%%%%%%%%%%%%%%%%%%%%%%%%%%%%%%%%%%%%%%%%%%%%%%%%%%%
%\FloatBarrier\clearpage
%\subsection{Groomed results}

The soft drop groomed \mjetsd/\pTjet distributions are shown in Fig.~\ref{fig:mjet_gr_WTAcomp} for both inclusive and WTA flavour-tagged \bquark jets, with normalisation set to results without a WTA flavour requirement. Comparisons between the ungroomed and groomed results are shown in Figs.~\ref{fig:mjet_grcomp_WTA} and \ref{fig:mjet_grcomp_noWTA} with and without WTA flavour tagging, respectively, with normalisation in both cases set to the ungroomed results. The shapes of the groomed distributions are affected by the removal of narrow, low-\mjet jets from the data sample, which are groomed away, pushing the peaks towards higher \mjet/\pTjet. The removal of soft, wide-angle radiation from surviving jets simultaneously makes the distributions more strongly peaked and shifted towards lower \mjet/\pTjet.
%The former effect tends to dominate at low \pTjet, while the latter is more significant at higher \pTjet.
The two-peak structure observed only at high \pTjet in the ungroomed case emerges at a lower threshold of $\pTjet > 20$\gev. This may suggest that grooming can be used as a tool to better separate quark-initiated \bquark-jets from gluon-initiated ones. Jets groomed near the kinematic threshold of $m_{\Bpm} / \pTjet$ exhibit a large peak, with a tail on the left from the broad \pTjet range.

Grooming slightly improves agreement between data and \pythia 8, especially in the case of WTA flavour tagging. This is expected since soft, wide-angle radiations are generally less well described perturbatively at fixed order than hard, collinear radiation. However, significant discrepancy remains in the tails of the distributions, with the data having more high-\mjet contributions than simulation, again pointing to a significant difference in high-virtuality direct versus in-shower \bquark-quark production.

%%%%%%%%%%%%%%%%%%%%%%%%%%%%%%%%%%%%%%%%%%%%%%%%%%%%%%%%%%%%%%%%%%%
\FloatBarrier
\section{Conclusion}

The \lhcb collaboration presents first measurements of \bquark-jet mass at forward rapidity in \pp collisions at $\s = 13$\tev. Winner-Take-All flavour tagging reduces the large-\mjet tail of the distributions by reducing gluon splitting contributions, offering a probe into in-shower production of \bquark quarks~\cite{Les_Houches_studies}. Comparisons to \pythia~8 predictions reveal significant deviations across the \mjet/\pTjet distributions with largest discrepancies in the tails of the distributions, which improve with WTA flavour tagging. These effects suggest that \bquark production models could be refined with improved parton shower prescriptions, such as including NRQCD heavy $q\bar{q}$ production~\cite{cooke2023_NRQCD}.

Soft drop grooming removes a substantial fraction of candidate \bquark jets and forms a secondary peak near the single-particle-jet kinematic limit. Agreement with \pythia 8 predictions improves with respect to ungroomed \mjet, especially at low \mjet/\pTjet where nonperturbative effects are reduced. These results will enable a comprehensive test of modern QCD calculations with heavy-flavour jets, for which a well-defined path to computation exists and calculations are in development~\cite{WTA_tagging, Larkoski:2023upz}.

Finally, these results are a baseline and proof-of-principle for future studies in heavy-ion collisions, where gluon splitting accesses the momentum transfer between hard probes and the quark-gluon plasma in a theoretically well-defined way.

%% file: acknowledgements.tex
\section*{Data availability}
The \lhcb experiment has agreed to the \cern Open Data policy that is summarised at
\href{https://opendata.cern.ch/docs/about}{https://opendata.cern.ch/docs/about}.
The \lhcb External Data Access Policy can be downloaded at
\href{https://open-data.cern.ch/record/410}{https://open-data.cern.ch/record/410}.
Data associated to the plots in this publication as well as in supplementary materials are made available on the \cern Document Server at
\href{https://cds.cern.ch/record/2932487}{https://cds.cern.ch/record/2932487}.
This manuscript has associated data in a HEPData repository at
\href{https://www.hepdata.net/record/ins2922449}{https://www.hepdata.net/record/ins2922449}.
An encapsulation of the
analysis is provided for measurements in the framework of Rivet at
\href{https://rivet.hepforge.org/analyses/LHCB\_2025\_I2922449.html}{https://rivet.hepforge.org/analyses/LHCB\_2025\_I2922449.html}.

\section*{Acknowledgements}
\noindent We gratefully acknowledge Arnd Behring, Simone Caletti, Andrew Larkoski, Simone Marzani, Rene Poncelet, and Daniel Reichelt for many useful discussions on flavour-tagging algorithms and on the jet mass.
%
% These Acknowledgements valid from 3-May-2019
%
%\noindent
We express our gratitude to our colleagues in the CERN
accelerator departments for the excellent performance of the LHC. We
thank the technical and administrative staff at the LHCb
institutes.
We acknowledge support from CERN and from the national agencies:
ARC (Australia);
CAPES, CNPq, FAPERJ and FINEP (Brazil); 
MOST and NSFC (China); 
CNRS/IN2P3 (France); 
BMBF, DFG and MPG (Germany); 
INFN (Italy); 
NWO (Netherlands); 
MNiSW and NCN (Poland); 
MCID/IFA (Romania); 
%MSHE (Russia); 
MICIU and AEI (Spain);
SNSF and SER (Switzerland); 
NASU (Ukraine); 
STFC (United Kingdom); 
DOE NP and NSF (USA).
%%%%%%%%%%%%%%%%%%%%%%%%%%%%%%%%%%%%%%%%%%%%%
We acknowledge the computing resources that are provided by ARDC (Australia), 
CBPF (Brazil),
CERN, 
IHEP and LZU (China),
IN2P3 (France), 
KIT and DESY (Germany), 
INFN (Italy), 
SURF (Netherlands),
Polish WLCG (Poland),
IFIN-HH (Romania), 
%RRCKI and Yandex LLC (Russia), 
PIC (Spain), CSCS (Switzerland), 
and GridPP (United Kingdom).
%%%%%%%%%%%%%%%%%%%%%%%%%%%%%%%%%%%%%%%%%%
We are indebted to the communities behind the multiple open-source
software packages on which we depend.
%%%%%%%%%%%%%%%%%%%%%%%%%%%%%%%%%%%%%%%%%%
Individual groups or members have received support from
%ARC and ARDC (Australia); % moved to national 16/01
Key Research Program of Frontier Sciences of CAS, CAS PIFI, CAS CCEPP, 
Fundamental Research Funds for the Central Universities,  and Sci. \& Tech. Program of Guangzhou (China);
Minciencias (Colombia);
EPLANET, Marie Sk\l{}odowska-Curie Actions, ERC and NextGenerationEU (European Union);
A*MIDEX, ANR, IPhU and Labex P2IO, and R\'{e}gion Auvergne-Rh\^{o}ne-Alpes (France);
%RFBR, RSF and Yandex LLC (Russia);
Alexander-von-Humboldt Foundation (Germany);
ICSC (Italy); 
%GVA, XuntaGal, GENCAT, Inditex, InTalent and Prog.~Atracci\'on Talento, CM (Spain);
Severo Ochoa and Mar\'ia de Maeztu Units of Excellence, GVA, XuntaGal, GENCAT, InTalent-Inditex and Prog. ~Atracci\'on Talento CM (Spain);
SRC (Sweden);
the Leverhulme Trust, the Royal Society and UKRI (United Kingdom).

%% file: supplementary-app.tex
\clearpage

\section*{Supplementary material for LHCb-PAPER-2025-009}
\label{sec:Supplementary-App}
\appendix

\section{Example mass distribution}
\label{app:mass_fit}

\begin{figure}[b!]
    \centering
    \includegraphics[width=0.8\linewidth]{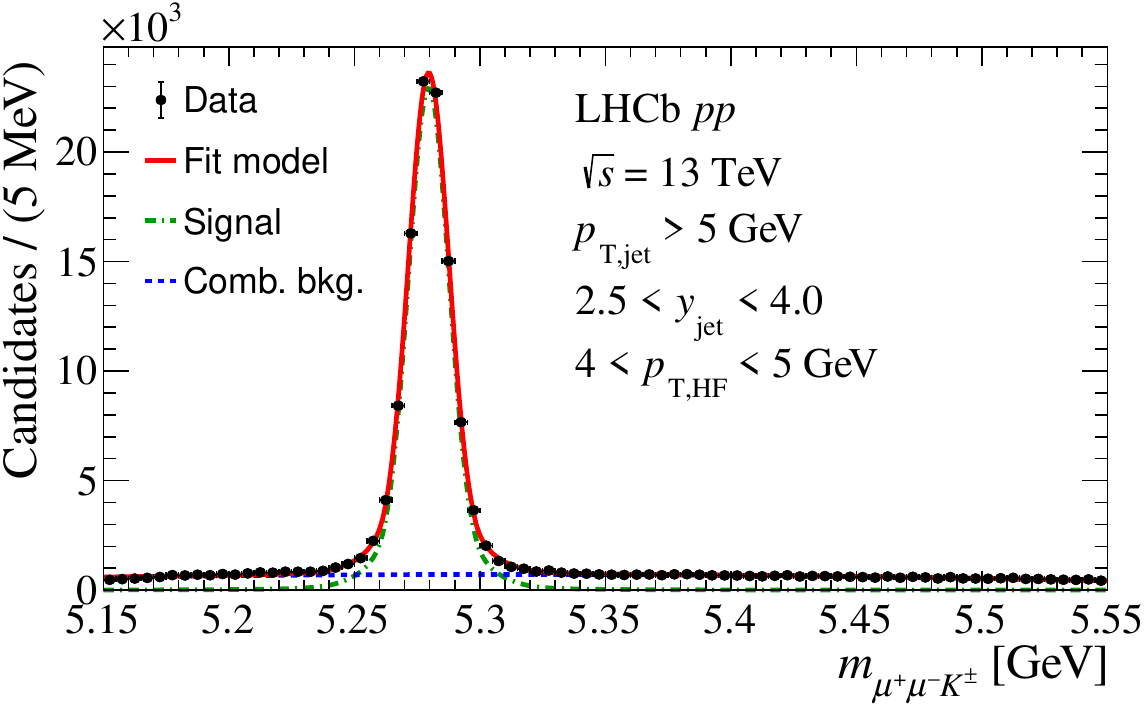}
    \caption{Distribution of the $\mup\mun\Kpm$ mass for candidate \Bpm mesons with transverse momentum \mbox{$4 < \pTHF < 5$\gev}. A parametric fit is shown in red, including contributions from signal, combinatorial background, and misidentified background.
    }
    \label{fig:example_massfit}
\end{figure}

An example of the \Bpm candidate mass fit described in the text is shown in Fig.~\ref{fig:example_massfit}.

%%%%%%%%%%%%%%%%%%%%%%%%%%%%%%%%%%%%%%%%%%%%%%%%%%%%%%%%%%%%%

\section{Systematic uncertainty summary}
\label{app:syst_uncert}

The approximate magnitude for each contribution to the systematic uncertainty on the \mjet/\pTjet distributions is summarized in Table~\ref{tab:total_systematic}.
This table reports approximate minima and maxima values over all distributions (including both WTA configurations, with and without grooming).
The dominant systematic uncertainty in this analysis is the uncertainty on the jet energy scale and resolution (JES/JER),
followed by the uncertainty on the unfolding procedure. These uncertainties tend to be larger in the sparse tails of the distributions, where small numerical variations correspond to large percentage variations.

\begin{table}[b!]
    \centering
    \caption{Summary of relative systematic uncertainties for the $B^{\pm}$-tagged \mjet/\pTjet distributions from various sources. These ranges include sparse bins at the edges of the distributions; the uncertainties tend to be smallest for the densely populated central values, and at lower \pTjet.}
    \begin{tabular}{c|c}  
        Source                  & Max uncertainty \\ \hline
        JES/JER                 & $50$\%           \\
        Unfolding               & $30$\%           \\
        Signal shape            & $14$\%           \\
        Sideband subtraction    & $6$\%            \\
        Track reconstruction efficiency        & $4$\%            \\
        PID efficiency          & $2$\%            \\
        Trigger efficiency            & $2$\%            \\ \hline  
        Total                   & $60$\%           \\ \hline 
    \end{tabular}
    \label{tab:total_systematic}
\end{table}

\clearpage

%% file: Authorship_LHCb-PAPER-2025-009.tex
% LHCb collaboration author list
% Data extracted on November 4th, 2025 at 1:52pm for paper reference LHCb-PAPER-2025-009
\centerline
{\large\bf LHCb collaboration}
\begin
{flushleft}
\small
R.~Aaij$^{38}$\lhcborcid{0000-0003-0533-1952},
A.S.W.~Abdelmotteleb$^{57}$\lhcborcid{0000-0001-7905-0542},
C.~Abellan~Beteta$^{51}$\lhcborcid{0009-0009-0869-6798},
F.~Abudin{\'e}n$^{57}$\lhcborcid{0000-0002-6737-3528},
T.~Ackernley$^{61}$\lhcborcid{0000-0002-5951-3498},
A. A. ~Adefisoye$^{69}$\lhcborcid{0000-0003-2448-1550},
B.~Adeva$^{47}$\lhcborcid{0000-0001-9756-3712},
M.~Adinolfi$^{55}$\lhcborcid{0000-0002-1326-1264},
P.~Adlarson$^{84}$\lhcborcid{0000-0001-6280-3851},
C.~Agapopoulou$^{14}$\lhcborcid{0000-0002-2368-0147}
% ...

{\footnotesize \it

% ...
$^{14}$Universit{\'e} Paris-Saclay, CNRS/IN2P3, IJCLab, Orsay, France\\
% ...
$^{38}$Nikhef National Institute for Subatomic Physics, Amsterdam, Netherlands\\
% ...
$^{47}$Instituto Galego de F{\'\i}sica de Altas Enerx{\'\i}as (IGFAE), Universidade de Santiago de Compostela, Santiago de Compostela, Spain\\
% ...
$^{51}$Physik-Institut, Universit{\"a}t Z{\"u}rich, Z{\"u}rich, Switzerland\\
% ...
$^{55}$H.H. Wills Physics Laboratory, University of Bristol, Bristol, United Kingdom\\
% ...
$^{57}$Department of Physics, University of Warwick, Coventry, United Kingdom\\
% ...
$^{60}$School of Physics and Astronomy, University of Glasgow, Glasgow, United Kingdom\\
$^{61}$Oliver Lodge Laboratory, University of Liverpool, Liverpool, United Kingdom\\
% ...
$^{69}$Syracuse University, Syracuse, NY, United States\\
% ...
$^{84}$Department of Physics and Astronomy, Uppsala University, Uppsala, Sweden, associated to $^{60}$\\
% ...
}
\end{flushleft}